\documentclass[journal=jctcce]{achemso}
\usepackage[utf8]{inputenc}
\usepackage{graphicx}
\usepackage{algorithm}
\usepackage{algpseudocode}
\usepackage{amsmath}
\usepackage{bm}
\usepackage{physics}          
\usepackage{mathtools}
\usepackage{multirow}
\usepackage{xr}        

\newcommand{\defeq}{\vcentcolon=}

\newcommand{\opr}[1]{\mathcal{#1}}

\newcommand{\vct}[1]{\bm{#1}}

\makeatletter
\def\BState{\State\hskip-\ALG@thistlm}
\makeatother
\title{Excited-state DMRG made simple with FEAST}

\author{Alberto Baiardi}
\affiliation{ETH Z\"{u}rich, Laboratorium f\"{u}r Physikalische Chemie, Vladimir-Prelog-Weg 2, 8093 Z\"{u}rich, Switzerland.}
\author{Anna Klára Kelemen}
\affiliation{ETH Z\"{u}rich, Laboratorium f\"{u}r Physikalische Chemie, Vladimir-Prelog-Weg 2, 8093 Z\"{u}rich, Switzerland.}
\author{Markus Reiher}
\affiliation{ETH Z\"{u}rich, Laboratorium f\"{u}r Physikalische Chemie, Vladimir-Prelog-Weg 2, 8093 Z\"{u}rich, Switzerland.}
\email{markus.reiher@phys.chem.ethz.ch}

\begin{document}

\begin{abstract}
\noindent We introduce DMRG[FEAST], a new method for optimizing excited-state many-body wave functions with the density matrix renormalization group (DMRG) algorithm.
Our approach applies the FEAST algorithm, originally designed for large-scale diagonalization problems, to matrix product state wave functions.
We show that DMRG[FEAST] enables the stable optimization of both low- and high-energy eigenstates, therefore overcoming the limitations of state-of-the-art excited-state DMRG algorithms.
We demonstrate the reliability of DMRG[FEAST] by calculating anharmonic vibrational excitation energies of molecules with up to 30 fully coupled degrees of freedom.
\end{abstract}

\maketitle

\section{Introduction}
\label{sec:intro}

Quantum-chemical formulations of the density matrix renormalization group (DMRG) algorithm\cite{White1992_DMRGBasis,White1993_DMRGBasis,Legeza2008_Review,Chan2008_Review,Zgid2009_Review,Marti2010_Review-DMRG,Schollwoeck2011_Review-DMRG,Chan2011_Review,Wouters2013_Review,Kurashige2014_Review,Olivares2015_DMRGInPractice,Szalay2015_Review,Yanai2015,Knecht2016_Chimia,Freitag2020_BookChapter,Baiardi2020_Review} have dramatically pushed the limits of full configuration interaction (full-CI) methods in the past two decades.
DMRG has opened up the possibility of calculating accurate wave functions for molecular systems including up to 100 particles and orbitals both for electronic\cite{Marti2008_DMRGMetalComplexes,Kurashige2013_MnCaCluster-DMRG,Yanai2014_WaterReduction-DMRG,Sharma2014_IronSulfur-DMRG,Stein2017_AutoCAS-IrCatalysis} and nuclear\cite{Oseledts2016,Baiardi2017_VDMRG,Baiardi2019_HighEnergy-vDMRG,Muolo2020_NEAP-DMRG,Feldmann2021_ProtonDensities,Glaser2021_BookChapter} problems.
Unlike approaches that leverage the sparsity of the CI coefficient tensor, such as the full-CI Quantum Monte Carlo\cite{Alavi2009_FCIQMCOriginal,Alavi2011_Initiator-FCIQMC} algorithm and selected CI methods,\cite{Malrieu1973_CIPSI-Original,HeadGordon2016_SelectedCI,Schriber2016_AdaptiveCI,Sharma2017_HBCI,Sharma2018_Fast-HBCI,Berkelbach2021_HBCI-Vibrational} the CI coefficient tensor is replaced with its low-rank tensor-train (TT) factorization\cite{Kolda2009_TensorReview,Oseledets2012_ALS} in DMRG.
The resulting wave function is known as matrix product state (MPS)\cite{McCulloch2007_FromMPStoDMRG} and can be optimized iteratively with the alternating least-squared (ALS) algorithm.\cite{Oseledets2012_ALS} 
The combination of MPS and ALS results in the DMRG method.
Whereas the quantum-chemical application of DMRG for ground-state many-body wave functions is well established, the design of efficient excited-state DMRG variants still represents a significant challenge.
In fact, the area law\cite{Hastings2007_AreaLaw}, which ensures that the ground-state wave function of short-ranged Hamiltonians can be encoded as compact MPSs, does not apply to excited states.
From a computational perspective, the ALS minimization algorithm relies on the variational principle and is therefore inherently tailored to ground-state calculations.
Finally, the number of excited states required for predicting, for instance, molecular spectra in a given energy range can be very large, especially for big systems and for high-energy regions.
An excited-state DMRG algorithm that is as efficient as ground-state DMRG and that can target many excited states in parallel is therefore still lacking.

We show, that the algorithm introduced in this work, namely DMRG[FEAST], can fill this gap.
DMRG[FEAST] extends the FEAST algorithm\cite{Polizzi2009_FEAST,Polizzi2014_FEAST-Iterative,Polizzi2015_FEAST-Zolotarev,Polizzi2016_FEAST-NonHermitian} to wave functions encoded as MPSs.
FEAST is a subspace diagonalization method for generalized eigenvalue problems in which the subspace basis is constructed by approximating the Hamiltonian Green function with a numerical integration of the underlying complex contour integral.
The subspace is calculated, in practice, by solving multiple independent linear systems defined in terms of the respective shift-and-invert Hamiltonian.\cite{Oseledets2012_ALS,Holz2012_ALSTheory,Kuprov2014_TensorTrain-NMR,Oseledts2016}
The present work combines for the first time FEAST and DMRG to design an algorithm that overcomes all the limitations of the existing state-of-the-art excited-state DMRG variants.
DMRG[FEAST] is capable of optimizing every state whose energy lies in a given energy interval, which is the only input parameter for the algorithm.
This is a major advantage compared to other state-targeting DMRG variants that either require a precise estimate of the excitation energies \texttt{a priori}, or rely on root-homing algorithms to follow the target state along with the optimization.\cite{Butscher1976_RootHoming,Baiardi2019_HighEnergy-vDMRG}

We illustrate the advantages of DMRG[FEAST] for the vibrational formulation of DMRG (vDMRG).\cite{Baiardi2017_VDMRG,Baiardi2019_HighEnergy-vDMRG}
Efficient and high-accuracy excited-state methods are particularly relevant for vibrational-structure calculations, where the calculation of anharmonic vibrational spectra with more than 10 atoms requires optimizing hundreds of excited states. 
We first benchmark the accuracy of DMRG[FEAST] by calculating the vibrational transition frequencies of ethylene.\cite{Delahaye2014_EthylenePES}
We demonstrate the reliability of DMRG[FEAST] computing the low-frequency anharmonic vibrations of uracil,\cite{Carrington2018_Uracil} a molecule of 30-modes that is hardly targeted by alternative vibrational CI-based schemes.

\section{DMRG for excited states}
\label{sec:theory}

We first briefly review DMRG theory, both for ground-\cite{Baiardi2017_VDMRG} and excited-\cite{Baiardi2019_HighEnergy-vDMRG} states, to introduce the basic notation.
We then introduce the DMRG[FEAST] algorithm which we apply to the vibrational Hamiltonian described at the end of this section.
However, we emphasize that the DMRG[FEAST] algorithm is completely general and can be applied to any Hamiltonian that can be encoded as a matrix product operator (MPO).

\subsection{Standard vDMRG}
\label{subsec:DMRG}

A full-CI wave function of an $L$-dimensional quantum system reads
\begin{equation}
  \ket{\Psi} = \sum_{\sigma_1\cdots\sigma_L}^{N_\text{max}} 
			    C_{\sigma_1\cdots\sigma_L} \ket{\sigma_1\cdots\sigma_L}
			 = \sum_{\bm{\sigma}}^{N_\text{max}} C_{\bm{\sigma}} \ket{\bm{\sigma}} \, ,
  \label{eq:CIwaveFunction}
\end{equation}
where the wave function $\ket{\Psi}$ is expressed as a linear combination of occupation number vectors (ONVs) $\ket{\sigma_1\cdots\sigma_L}$ containing the single-state occupations $\sigma_i$.
Eq.~(\ref{eq:CIwaveFunction}) holds for both electronic and vibrational wave functions, with the respective ONVs defined for an appropriate one-particle basis of up to $L$ one-particle states.
In DMRG, the CI tensor $C_{\sigma_1\cdots\sigma_L}$ is approximated in the TT format\cite{Oseledets2012_ALS},

\begin{equation}
  C_{\sigma_1\cdots\sigma_L}
    = \sum_{\sigma_1\cdots\sigma_L} \sum_{a_1, \dots, a_{L-1}}^m 
      M_{1, a_1}^{\sigma_1} M_{a_1, a_2}^{\sigma_2} \cdots M_{a_{L-1}, 1}^{\sigma_{L}} \, .
  \label{eq:MPS}
\end{equation}
which yields the so-called matrix product state (MPS) wave function \textit{ansatz}.\cite{Rommer1997_MPS-Ansatz}. In 
Eq.~(\ref{eq:MPS}) the CI tensor is expressed as a product of $L$ tensors $\{M_{a_{i-1},a_i}^{\sigma_i}\}$ that are contracted along the auxiliary indices $a_i$.
The maximum value of the auxiliary indices, called the bond dimension $m$, plays a pivotal role in DMRG.
Although an arbitrary full CI wavefunction is represented exactly as an MPS with a bond dimension that grows exponentially with $L$, low values of $m$  are in practice often sufficient for encoding the ground state of vibrational\cite{Oseledts2016,Baiardi2017_VDMRG,Baiardi2019_HighEnergy-vDMRG} and electronic\cite{Chan2002_DMRG,Legeza2003_DMRG-LiF,Marti2008_DMRGMetalComplexes} molecular Hamiltonians. 

\noindent Analogously, a single-state-driven decomposition of an arbitrary Hamiltonian can be obtained as an MPO:\cite{Schollwoeck2011_Review-DMRG, McCulloch2007_FromMPStoDMRG}
\begin{equation}
  \opr{H} = \sum_{\vct{\sigma},\vct{\sigma'} } \sum_{b_1, \dots, b_{L-1}}^{\vct{r}} 
            H_{1, b_1}^{\sigma_1, \sigma'_1} \cdots H_{b_{L-1},1}^{\sigma_L,\sigma'_L}
            \ket{\vct{\sigma}}\bra{\vct{\sigma'}} \, ,
  \label{eq:MPO}
\end{equation}
While the MPS approximates the CI wave function for a fixed value of $m$, the Hamiltonian is represented exactly by Eq.~\eqref{eq:MPO}.
The sweep-based optimization strategy of DMRG optimizes the MPS representation of the ground state wave function in an iterative fashion.
In the single-site variant of DMRG, the energy expectation value is minimized with respect to variation of the $M_{a_{i-1}, a_i}^{\sigma_i*}$ tensor.
This results in an eigenvalue problem for tensor $\bm{M}^{\sigma_i}$,

\begin{equation}
  \sum_{b_{i-1},b_i}\sum_{a_{i-1}'} \sum_{a_i'} \sum_{\sigma_i'}
    L_{a_{i-1},a_{i-1}'}^{b_{i-1}} H_{b_{i-1},b_i}^{\sigma_i,\sigma_i'} R_{a_i,a_i'}^{b_i} M_{a_{i-1}',a_i'}^{\sigma_i'} 
  = E M_{a_{i-1},a_i}^{\sigma_i} \, ,
  \label{eq:EigenValue}
\end{equation}
where $L_{a_{i-1},a'_{i-1}}^{b_{i-1}}$ and $R_{a_i, a'_i}^{b_i}$ are the so-called boundaries\cite{Schollwoeck2011_Review-DMRG,Keller2015_MPS-MPO-SQHamiltonian} that collect the partial contraction between the MPS and the MPO for all sites smaller and larger than $i$, respectively.
In single-site DMRG, Eq.~(\ref{eq:EigenValue}) is solved one site at a time (in one microiteration step), by increasing $i$ from 1 to $L$ (forward sweep).
The optimization is then repeated in the opposite direction, starting from $i$=$L$ (backward sweep).
At each microiteration step, Eq.~(\ref{eq:EigenValue}) is solved and the MPS is updated by selecting the lowest energy root. 
The sweep optimization may converge to a local minimum of the energy functional, especially in the presence of symmetries in the Hamiltonian.
This issue can be addressed with the so-called subspace expansion algorithm\cite{McCulloch2015_Mixing} or by optimizing two consecutive sites simultaneously, which defines the two-site variant of DMRG.\cite{Keller2015_MPS-MPO-SQHamiltonian}

\noindent Several algorithms have been proposed for extending DMRG to excited states.\cite{Dorando2007_TargetingExcitedStates,Devakul2017,Yu2017_ShiftAndInvertMPS,Baiardi2019_HighEnergy-vDMRG}
The lowest-energy eigenstate can be targeted with conventional DMRG by constraining the optimization to the variational space orthogonal to the ground-state wave function.\cite{McCulloch2007_FromMPStoDMRG,Keller2015_MPS-MPO-SQHamiltonian} 
This procedure can be straightforwardly extended to higher-energy states but since it is inherently sequential, it becomes very inefficient for high-energy states.
A more appealing alternative is to apply DMRG to the shift-and-inverted (S\&I) counterpart of Eq.~(\ref{eq:EigenValue}).
The resulting method, DMRG[S\&I]\cite{Baiardi2019_HighEnergy-vDMRG}, is more efficient than standard, state-specific DMRG.
However, its numerical stability heavily depends on the choice of the shift parameter $\eta$ and on the root selected at each microiteration step to propagate the boundaries.
Root-homing techniques\cite{Butscher1976_RootHoming,Devakul2017,Baiardi2019_HighEnergy-vDMRG} were shown to enhance the efficiency of DMRG[S\&I], but the optimization remains challenging for high-energy states.

\subsection{DMRG[IP]}
\label{subsec:IP}

The DMRG[IP] method combines the Inverse Power (IP) algorithm and DMRG into a method in which, similarly to DMRG[S\&I], a single excited state is targeted.
However, as opposed to the S\&I technique, DMRG[IP] does not require the selection of a root in each microiteration step.
Starting from an initial wave function guess $\ket{\Psi_{0}}$, in each iteration step $k$ of the IP algorithm the S\&I operator

\begin{equation}
  \Omega_\eta = \left( \mathcal{H} - \eta \mathcal{I} \right)
  \label{eq:SandI_Operator}
\end{equation}
is applied to the wave function calculated at the $k$-th iteration step as

\begin{equation}
  \ket{\Psi_{k}} := ( \opr{H} - \eta \opr{I})^{-1} \ket{\Psi_{k-1}}\quad\quad k=1, \dots, N_{iter} \, .
  \label{eq:invPI}
\end{equation}
The wave function converges to the eigenfunction with energy closest to the shift $\eta$ for $k \rightarrow +\infty$.
In practice, the following equation is solved:

\begin{equation}
    \Gamma_\eta \ket{\Psi_{k}} = \ket{\Psi_{k-1}} \, ,
  \label{eq:IPI_sol}
\end{equation}
with
\begin{equation}
                \ket{\Psi_{k}} := \frac{\ket{\Psi_{k}}}{\norm{\Psi_{k}}}\, ,
  \label{eq:IPI_solb}
\end{equation}
where $\Gamma_\eta = \Omega_\eta^{-1} = ( \opr{H} - \eta \opr{I}) $.
The solution of Eq.~\eqref{eq:IPI_sol}, $\ket{\Psi_{k}}$, defines the right-hand side in the next IP iteration (referred to in the following as ``IP macroiteration'').
The convergence rate of the IP method depends on the following ratio:\cite{Saad2011_Book}
\begin{equation}
  \rho =  \abs{\frac{E^{(n)} - \eta}{E^{(n+1)} - \eta}} \, ,
  \label{eq:ConvergenceRateIPI}
\end{equation}
where $E^{(n)}$ is the target eigenvalue and $E^{(n+1)}$ is the second eigenvalue closest to the shift $\eta$.
Therefore, fast convergence will be ensured if the shift is chosen close to an exact eigenvalue of $\opr{H}$. 

\noindent In DMRG[IP], $\Gamma_\eta$ is encoded as an MPO and $\ket{\Psi_{k}}$ is approximated, in each iteration step, as an MPS with a fixed bond dimension $m$.
Since the application of the MPO onto an MPS increases its bond dimension,\cite{Schollwoeck2011_Review-DMRG} the solution to Eq.~\eqref{eq:IPI_sol} with DMRG can only be approximate.
Following Ref.~\citenum{Oseledts2016}, we approximate the solution to Eq.~(\ref{eq:IPI_sol}) with the least-squares method.
The optimal MPS of a given bond dimension $m$ is then defined by the minimum of the functional $O[\Psi_{k}]$\cite{Oseledets2012_ALS,Holz2012_ALSTheory},

\begin{equation}
 \begin{aligned}
  O_\eta \left[ \Psi_{k} \right] &= 
    \left\| \Gamma_\eta \ket{\Psi_{k}} - \ket{\Psi_{k-1}} \right\|^2 \\
      &= \langle \Psi_{k} | \Gamma_\eta^2 | \Psi_{k} \rangle 
       - 2 \Re \langle \Psi_{k}| \Gamma_\eta | \Psi_{k-1} \rangle
       + \langle \Psi_{k-1}| \Psi_{k-1} \rangle
 \end{aligned}
 \label{eq:residualMinimum}
\end{equation}
that is calculated with respect to the variation of the MPS site tensors, that is, $\text{argmin}_{M_{a_{i-1},a_i}^{\sigma_i,(k-1)}}(O_\eta \left[ \Psi_{k} \right])$.
For $\eta$ values smaller than the smallest eigenvalues of $\mathcal{H}$, $\Gamma_\eta$ is positive definite and the minimum of Eq.~\eqref{eq:residualMinimum} is also the minimum of the functional $\tilde{O}_\eta$ defined as:
\begin{equation}
  \tilde{O}_\eta  \left[ \Psi_{k} \right] = \langle \Psi_{k} | \Gamma_\eta | \Psi_{k} \rangle 
       - 2 \langle \Psi_{k-1} | \Psi_{k} \rangle\, .
 \label{eq:residualMinimum_Equivalent}
\end{equation}

$\Gamma_\eta$ is not positive definite otherwise and the minima of Eq.~\eqref{eq:residualMinimum} and \eqref{eq:residualMinimum_Equivalent} do not necessarily coincide.
However, Rhakuba and Oseledets argued\cite{Oseledts2016} that the equivalence holds in practice for any choice of $\eta$, provided that the initial guess for the solution of the linear system is a good approximation of the targeted eigenstate.
In DMRG[IP], the functional defined in Eq.~(\ref{eq:residualMinimum_Equivalent}) is minimized with the ALS algorithm, one site tensor at a time. This is in contrast to standard DMRG, where the variational minimum is defined with respect to the energy expectation value.\cite{Schollwoeck2011_Review-DMRG}
By denoting the tensor entering the MPS parametrization of the state $\ket{\Psi_{k-1}}$ for site $i$ as $M_{a_{i-1},a_i}^{\sigma_i,(k-1)}$, and the tensor associated with $\ket{\Psi_{k}}$ as $M_{a_{i-1},a_i}^{\sigma_i(k)}$, the first term in Eq.~(\ref{eq:residualMinimum_Equivalent}) can be written, for site $i$, as
\begin{equation}
  \mel{\Psi_{k}}{\Gamma_\eta}{\Psi_{k}} 
    = \sum_{\sigma_i, \sigma_i'} \sum_{\substack{a_{i-1} a_i \\ a_{i-1}' a_i'}} \sum_{b_{i-1},b_i}
      M_{a_{i-1}, a_i}^{\sigma_i,(k)} L_{a_{i-1},a'_{i-1}}^{b_{i-1}} G^{\sigma_i, \sigma_i'}_{b_{i-1}, b_i} 
      R_{a_i, a_i'}^{b_i} M_{a_{i-1}', a_i'}^{\sigma_i',(k)} \, ,
  \label{eq:FirstContraction}
\end{equation}
where $G^{\sigma_i, \sigma_i'}_{b_{i-1}, b_i}$ are the tensors defining the MPO representation of $\Gamma_\eta$.
The tensors $L_{a_{i-1},a_{i-1}'}^{b_{i-1}}$ and $R_{a_i, a_i'}^{b_i}$ are the boundaries\cite{Schollwoeck2011_Review-DMRG,Keller2015_MPS-MPO-SQHamiltonian} introduced above.
The second term in Eq.~(\ref{eq:residualMinimum_Equivalent}) can be expressed as
\begin{equation}
  \langle \Psi_{k} | \Psi_{k-1} \rangle = \sum_{\sigma_i} \sum_{a_{i-1}, a_i} \sum_{a_{i-1}', a_i'} 
     A_{a_{i-1},a_{i-1}'} M_{a_{i-1},a_i}^{\sigma_i,(k)} M_{a_{i-1}',a_i'}^{\sigma_i,(k-1)} B_{a_i',a_i} \, ,
  \label{eq:SecondContraction}
\end{equation}
where the tensors $A_{a_{i-1},a_{i-1}'}$ and $B_{a_i',a_i}$ collect the partial contractions of the MPSs $\ket{\Psi_{k-1}}$ and $\ket{\Psi_{k}}$ for sites smaller and larger than $i$, respectively.
By exploiting Eqs.~(\ref{eq:FirstContraction}) and (\ref{eq:SecondContraction}), 
the tensor $\tilde{M}_{a_{l-1},a_l}^{\sigma_l}$ minimizing Eq.~(\ref{eq:residualMinimum_Equivalent}) is obtained from
\begin{equation}
  \sum_{\sigma_i'} \sum_{a_{i-1}', a_i'} \sum_{b_{i-1},b_i} 
    L^{a_{i-1},a_{i-1}'}_{b_{i-1}} G^{\sigma_i, \sigma_i'}_{b_{i-1}, b_i} R_{b_i}^{a_i, a_i'}
    \tilde{M}_{a_{i-1}', a'_i}^{\sigma'_l} 
  = \sum_{a_{i-1}',a_i'} A_{a_{i-1},a_{i-1}'} \tilde{M}^{\sigma_l}_{\tilde{a}_{i-1},\tilde{a}_i} 
                         B_{ \tilde{a}_l, a_l} \, .
  \label{eq:ContractedLinearSystem}
\end{equation}

Eq.~\eqref{eq:ContractedLinearSystem} can be recast as a linear system of equations that can be solved with the Generalized Minimal Residual (GMRES) method\cite{Saad1986_GMRES,Koch2000_Bratwurst} that solves Eq.~\eqref{eq:ContractedLinearSystem} in the Krylov space generated by the powers of the $\Gamma_\eta$ operator.
Note that, in analogy to standard DMRG, the functional defined in Eq.~(\ref{eq:residualMinimum_Equivalent}) can be minimized with respect to the variation of two neighboring tensors, as is done in the two-site DMRG.\cite{Schollwoeck2011_Review-DMRG,Keller2015_MPS-MPO-SQHamiltonian}

\noindent Following Ref.~\citenum{Oseledts2016}, we do not run the DMRG sweeps for solving Eq.~(\ref{eq:ContractedLinearSystem}) until convergence, but rather for a fixed number of sweeps $N_\text{sweeps}$.
Therefore, we construct the right-hand side of the $k$-th IP iteration in Eq.~\eqref{eq:IPI_sol} from the partially optimized MPS at the ($k$-1)-th iteration and repeat the procedure until convergence, which is monitored through the parameter $\lambda$ defined as:

\begin{equation} 
  \lambda= \left| \langle \Psi_{k} | \Psi_{k-1} \right \rangle|^2 \, .
  \label{eq:convergenceThreshold}
\end{equation}

\noindent The solution of Eq.~(\ref{eq:ContractedLinearSystem}) is unique, and therefore, DMRG[IP] does not require a root-homing algorithm to select the proper eigenfunction after each microiteration step.
However, as for DMRG[S\&I], choosing the appropriate $\eta$ value to target a given eigenstate remains challenging for high-energy states.
In the following section, we show that this limitation can be lifted with FEAST.

\subsection{DMRG[FEAST]}
\label{subsec:FEAST}

The FEAST algorithm\cite{Polizzi2009_FEAST,Galgon2011_FEAST-Review,Polizzi2014_FEAST-Iterative,Polizzi2015_FEAST-Zolotarev,Polizzi2016_FEAST-NonHermitian} is an iterative subspace diagonalization method for generalized eigenvalue problems.
Unlike Davidson-type algorithms,\cite{Davidson1975_DavidsonDiagonalization,Sleijpen2000_JacobiDavidson} the subspace dimension in FEAST is fixed and the basis in which the diagonalization is carried out is generated by solving a contour integral in the complex plane, for a set of guess vectors.
We first introduce the method as applied to the standard eigenvalue problem and, then, introduce DMRG[FEAST].

\noindent Let $I_E = [E_\text{min}, E_\text{max}]$ be an energy interval that contains $M$ eigenvalues of $\mathcal{H}$ with the corresponding eigenfunctions $I_M = \{\Psi^{(1)}, \dots,\Psi^{(M)}\}$.
We consider a closed curve $\mathcal{C}$ in the complex plane enclosing $I_E$.
The projector $\mathcal{P}_M$ associated with $I_M$ can be expressed as a contour integral\cite{Polizzi2009_FEAST} based on Cauchy's integral theorem\cite{Galgon2011_FEAST-Review}
\begin{equation}
  \mathcal{P}_{M} = \sum_{j=1}^{M} \ket{\Psi^{(j)}} \bra{\Psi^{(j)}}
                    = \frac{1}{2\pi \mathrm{i}} \oint_{\mathcal{C}}
                      (z \opr{I} - \opr{H})^{-1} dz \, .
  \label{eq:FEAST_Projector}
\end{equation}
We apply $\mathcal{P}_{M}$ onto a set of $N$ linearly independent guess vectors $\{\Phi_\text{guess}^{(1)}, \ldots, \Phi_\text{guess}^{(N)}\}$ 
\begin{equation}
  S_N \defeq 
    \{ \mathcal{P}_{M} \Phi_\text{guess}^{(1)}, \ldots, \mathcal{P}_{M} \Phi_\text{guess}^{(N)} \}\, .
  \label{eq:ProjectedSet}
\end{equation}

\noindent Eq.~\eqref{eq:ProjectedSet} corresponds to the subspace basis in which $\mathcal{H}$ is diagonalized in the FEAST algorithm.
If $N$ is equal to the number of eigenstates included in $I_E$,  the eigenpairs with eigenvalues within the interval $I_E$ can be obtained.
We note that, in practice, overestimating the subspace size by providing $N>M$ guess vectors is necessary to ensure convergence.\cite{Polizzi2009_FEAST}
This overestimation however, leads to spurious eigenvalues that do not correspond to those of the full Hamiltonian.
If the number of guess vectors is $N<M$, the basis $S_N$ does not span the appropriate subspace and the diagonalization does not yield the exact eigenvalues lying in the target interval.
Calculating the subspace basis with Eq.~(\ref{eq:FEAST_Projector}) would require inverting the $(z \opr{I} - \opr{H})$ operator and the exact evaluation of the contour integral.
The key idea of FEAST is to approximate the complex contour integral with an $N_\text{p}$-point numerical quadrature.
Each element of the $S_N$ set is therefore expressed as

\begin{equation}
  \mathcal{P}_{M} \Phi^{(i)}_{\text{guess}}
    \approx \frac{1}{2\pi \mathrm{i} } \sum_{k=1}^{N_p} \omega_k (z_k \opr{I} - \opr{H})^{-1} \Phi^{(i)}_{\text{guess}}
    \defeq \sum_{k=1}^{N_p} \omega_k \Phi^{(i,k)} \, ,
  \label{eq:ProjectorQuadrature}
\end{equation}
where $\omega_k$ and $z_k$ are nodes and weights defined by the numerical quadrature.
The wave function $\Phi^{(i,k)}$ associated with a given quadrature node $k$ and guess $i$ is obtained as solution of the following linear system:

\begin{equation}
  (z_k \opr{I}- \opr{H}) \Phi^{(i,k)} = \Phi^{(i)}_{\text{guess}} \, ,
  \label{eq:linfeast}
\end{equation}
where $(z_k \opr{I}- \opr{H}) = -\Omega_{z_k}$, with $\Omega_{z_k}$ being the S\&I operator defined in Eq.~(\ref{eq:SandI_Operator}).
Note that the linear systems defined by Eq.~(\ref{eq:linfeast}) are mutually independent, which makes the FEAST algorithm trivially parallelizable.
Each element of $S_N$ is then obtained by evaluating the summation over the quadrature nodes. If Eq.~\eqref{eq:linfeast} is solved exactly, the only approximation of FEAST is the numeric integration of Eq.~\eqref{eq:ProjectorQuadrature}. 
In this case FEAST returns the exact eigenpairs in a single iteration for a large enough quadrature grid.

\noindent In DMRG[FEAST], we express the initial guess vectors $\{\Phi^{(1)}_{\text{guess}}, \ldots, \Phi^{(M)}_{\text{guess}}\}$ as a set of MPSs, each with bond dimension $m$ and $\Omega_{z_k}$ associated with each quadrature point in Eq.\eqref{eq:linfeast} as an MPO.
The solution of Eq.~\eqref{eq:linfeast}, $\Phi^{(i,k)}$, is approximated with the minimum defined by Eq.~\eqref{eq:ContractedLinearSystem} for an MPS with a fixed bond dimension $m$.
The FEAST subspace basis is obtained by optimizing this MPS with the iterative scheme 
introduced for DMRG[IP]. Note that the operator appearing in Eq.~(\ref{eq:linfeast}) 
was introduced for FEAST and has the opposite sign compared to the S\&I operator of the DMRG[IP] 
algorithm. Therefore, the MPO must be scaled by an overall factor of 
-1 before solving Eq.~(\ref{eq:linfeast}).

Because the sum of two MPSs with bond dimension $m$ is represented by an MPS with bond dimension $2m$,\cite{Oseledts2016,Guo2018_DMRG-Hylleraas} Eq.~\eqref{eq:ProjectorQuadrature} yields a subspace composed by MPSs with bond dimension $M \times m$.

We therefore expect the MPS representation of the target wave functions associated with $I_E$ to be more accurate than for the original guess states.
We report a pseudocode of the DMRG[FEAST] algorithm in Section~\ref{sec:computationalDetails}.
Note that for Hermitian Hamiltonians Eq.\eqref{eq:ProjectorQuadrature} can be rewritten as\cite{Polizzi2009_FEAST}

\begin{equation}
  \begin{aligned}
    \mathcal{P}_{M} \Phi^{(i)}_{\text{guess}} &= 
    \frac{1}{2\pi \mathrm{i}} \int_{\mathcal{C}^+} \left[ (z \opr{I} - \opr{H})^{-1} \right] \Phi^{(i)}_{\text{guess}}  \text{d} z 
                            + \frac{1}{2\pi \mathrm{i}}\int_{\mathcal{C}^-} \left[ (z \opr{I} - \opr{H})^{-1} \right] \Phi^{(i)}_{\text{guess}}  \text{d} z \\
    &= \frac{1}{2} \int_0^1 \left[ (z(\theta) \, \opr{I} - \opr{H})^{-1}\Phi^{(i)}_{\text{guess}} \right] \bar{r} e^{i\pi\theta} \text{d} \theta 
   - \frac{1}{2} \int_1^0 \left[ (z(\theta)^\star \opr{I} - \opr{H})^{-1}\Phi^{(i_{\text{guess}})} \right] \bar{r} e^{-i\pi\theta} \text{d} \theta \\
    &= \frac{1}{2} \int_0^1 \left[ (z(\theta) \, \opr{I} - \opr{H})^{-1} \right] \Phi^{(i)}_{\text{guess}} \bar{r} e^{i\pi\theta} \text{d} \theta 
   + \frac{1}{2} \int_0^1 \left[ (z(\theta) \opr{I} - \opr{H})^{-1} \right]^\dagger \Phi^{(i)}_{\text{guess}} \bar{r} e^{-i\pi\theta} \text{d} \theta \\
    &= \frac{1}{2} \int_0^1 \left( \Phi^{(i,z)} + \Phi^{(i,z),\star} \right) \text{d} \theta 
  \end{aligned}
  \label{eq:ProjectorHermitian}
\end{equation}
where $\mathcal{C}^+$ is the part of the circle lying in the upper positive complex plane, and

\begin{equation}
  z(\theta) = \frac{E_\text{max}+E_\text{min}}{2} + \left( \frac{E_\text{max}-E_\text{min}}{2} \right) e^{2\pi\text{i}\theta} \, .
  \label{eq:ZetaDef}
\end{equation}
\noindent Eq.~\eqref{eq:ProjectorHermitian} expresses $\mathcal{P}_{N} \Phi^{(i)}_{\text{guess}}$ as a complex integral over the positive half of the complex circle $\mathcal{C}$.
It follows, that for achieving the same accuracy as for a non-Hermitian operator, only half as many quadrature points are needed.
In the original FEAST algorithm,\cite{Polizzi2009_FEAST} Eq.~\ref{eq:ProjectorHermitian} is further simplified based on the relation $\left( \Phi^{(i, z)} + \Phi^{(i,z),\star} \right) = 2 \Re(\Phi^{(i,z)})$ that, however, does not hold for MPSs.

\noindent Compared to the original FEAST algorithm, DMRG[FEAST] introduces two additional approximations.
First, the eigenfunctions of the Hamiltonian are expressed as a linear combination of MPSs with a fixed bond dimension $m$ and, therefore, the energy convergence with respect to the bond dimension must be monitored.
Second, we approximate the solutions of Eq.~\eqref{eq:linfeast} by running the sweep optimization for a finite number of sweeps ($N_\text{sweeps}$) and GMRES iteration ($N_\text{GMRES}$) and, therefore, approximate the $S_N$ basis.
We balance this inaccuracy by repeating the FEAST iteration step for $N_\text{FEAST}$ iterations.
This approach is advantageous because Eq.~\eqref{eq:linfeast} may be ill-conditioned, especially in energy ranges with a high density of states where a node $z_k$ may be close to an eigenvalue of $\mathcal{H}$. \\
We highlight that the number of eigenstates $M$ lying in the interval $I_E$ is not known \textit{a priori}.
DMRG[FEAST] must, therefore, be repeated for increasing $N_\text{guess}$ values to ensure that all roots are converged.

\noindent We assess the convergence of DMRG[FEAST] based on three different criteria.
A single microiteration step of a DMRG sweep is converged if the relative residual of the local linear system (Eq.~\eqref{eq:ContractedLinearSystem}) falls below a given threshold $\eta_\text{micro}$ before $N_\text{GMRES}$ iterations

\begin{equation}
  \frac{\textbf{F}\vct{x} - \vct{c}}{\textbf{F}\vct{x}_0 - \vct{c}} < \eta_\text{micro}\, .
  \label{eq:FirstConvergence}
\end{equation}
where $\vct{x}$ is a vector that collects the entries of the $\tilde{M}_{a_{i-1}',a_i'}^{\sigma_i'}$ tensor, $\vct{c}$ collects the entries of the right-hand side Eq.~\ref{eq:ContractedLinearSystem}, and $\textbf{F}$ is defined as:

\begin{equation}
  F_{(a_{i-1}'a_i'\sigma_i'),(a_{i-1}a_i\sigma_i)} = 
  \sum_{b_{i-1},b_i} 
    L^{a_{i-1},a_{i-1}'}_{b_{i-1}} G^{\sigma_i, \sigma_i'}_{b_{i-1}, b_i} R_{b_i}^{a_i, a_i'} \, .
  \label{eq:ContractedLinearSystemOperator}
\end{equation}

We monitor the convergence of the sweep-based solution of the linear system based on the overlap of the MPS at the current sweep with the MPS obtained at the end of the previous sweep.
When this overlap falls below a second threshold $\eta_\text{overlap}$, \textit{i.e.}

\begin{equation}
  \text{Re} \left( \abs{\bra{\Psi^k}\ket{\Psi^{k-1}}} \right) < \eta_\text{overlap}
  \label{eq:SecondConvergence}
\end{equation}
the MPS is considered stationary and therefore the sweeping procedure is stopped before reaching $N_\text{sweeps}$.
The convergence of DMRG[FEAST] is assessed following the original implementation of FEAST.\cite{Polizzi2009_FEAST}
Denoting the eigenvalues at the $i$-th FEAST iteration as $E^{(i)}$, and $E^{(i+1)}$ the same quantity at the ($i$+1)-th iteration, the FEAST iterations are stopped if:

\begin{equation}
 \frac{\sum_j \left| E_j^{(i)} - E_j^{(i+1)} \right|}{\sum_j E_j^{(i)}} < \eta_\text{FEAST}
  \label{eq:EigenvaluesRatio}
\end{equation}
$\eta_\text{FEAST}$ being a third convergence threshold.

\noindent In conclusion, we highlight the advantages of DMRG[FEAST] over DMRG[IP].
The basis of the subspace diagonalization in DMRG[FEAST] is obtained as the target eigenfunctions are obtained in DMRG[IP] for a single iteration.
However, the shift in Eq.~\eqref{eq:linfeast} is determined by the quadrature defined on the user-defined interval, as opposed to the target energy in Eq.~\eqref{eq:IPI_sol} for DMRG[IP].
This is advantageous, since the convergence rate of DMRG[IP] depends on the proximity of the shift to the target state.
As we will show in the next section, the selection of $I_E$ is not critical because the interval must merely be sufficiently large to include all target eigenfunctions.
As opposed to iteratively improving a single guess by the repeated applications of the shifted operator and, therefore, obtaining the excited state in a sequential fashion as in DMRG[IP], in DMRG[FEAST] multiple independent linear systems corresponding to different quadrature nodes and guesses are solved.
Therefore, DMRG[FEAST] can be trivially parallelized at different levels.
First, multiple energy regions can be targeted simultaneously with different DMRG[FEAST] runs.
Within each run, all linear systems can be solved independently of the others.
Finally, the solution of the linear system with DMRG can be parallelized as for conventional DMRG.\cite{Brabec2021_Parallel-DMRG}

\subsection{Vibrational DMRG}
\label{subsec:vDMRG}

DMRG[FEAST] can be applied to any Hamiltonian that can be encoded as an MPO.
This is the case for electronic,\cite{Keller2015_MPS-MPO-SQHamiltonian,Chan2016_MPO-MPS,Battaglia2018_RelativisticDMRG} vibrational,\cite{Baiardi2017_VDMRG,Muolo2020_NEAP-DMRG}, and rotational\cite{Roy2018_RotationalDMRG} Hamiltonians.
In this work, we apply DMRG[FEAST] to the vibrational formulation of DMRG, namely vDMRG.\cite{Baiardi2017_VDMRG}

\noindent We obtain the MPO representation of the vibrational Hamiltonian of an $L$-mode system starting from the so-called Watson Hamiltonian expressed in terms of Cartesian normal modes ($\bm{Q}$):\cite{Watson1968_Hamiltonian,Papousek}

\begin{equation}
  \mathcal{H}_\text{vib} = - \sum_{i=1}^L \frac{\partial^2}{\partial Q_i^2} 
                           + \mathcal{V}(Q_1, \ldots, Q_L)
  \label{eq:WatsonHam}
\end{equation}
and express the potential operator $\mathcal{V}(Q_1, \ldots, Q_L)$ with a Taylor expansion around a stationary point of the potential energy surface (PES):

\begin{equation}
  \mathcal{V}(Q_1, \ldots, Q_L) = \sum_{i=1}^L \omega_i^2 Q_i^2 
                                + \frac{1}{6} \sum_{ijk} k_{ijk} Q_i Q_j Q_k
                                + \frac{1}{24} \sum_{ijkl} k_{ijkl} Q_i Q_j Q_k Q_l\, .
  \label{eq:PotTaylor}
\end{equation}

\noindent The MPO representation of Eq.~\ref{eq:WatsonHam} based on the potential defined in Eq.~\ref{eq:PotTaylor} is obtained with the algorithm presented in Ref.~\citenum{Baiardi2017_VDMRG}.
We encode $\mathcal{H}_\text{vib}$ in the MPO format by expressing the position operator $Q_i$ and its conjugate momentum $P_i$ in terms of the canonical bosonic second-quantization operators $b_i^\dagger, b_i$ as follows:

\begin{equation}
  \begin{aligned}
    Q_i &= \frac{1}{\sqrt{2}} \left( b_i^\dagger + b_i \right) \\
    P_i &= \frac{\mathrm{i}}{\sqrt{2}} \left( b_i^\dagger - b_i \right) \\
  \end{aligned}
  \label{eq:CanonicalQuantization}
\end{equation}

\noindent It follows from Eq.~\ref{eq:CanonicalQuantization} that each site $i$ of the DMRG lattice corresponds to a given mode.
Harmonic modes fulfill the Bose-Einstein statistics and, therefore, the local dimension of each site ($N_\text{max}$ in Eq.~\ref{eq:MPS}) is unbounded.
In practice, we set $N_\text{max}$=6 for all calculations since higher energy ONVs give negligible contributions to the final MPS wave function.

\section{Computational Details}
\label{sec:computationalDetails}

We outline the DMRG[IP] and DMRG[FEAST] algorithms in Algorithm~\ref{alg:DMRG_IP} and \ref{alg:DMRG_FEAST}, respectively. 
\begin{algorithm}[htbp!]
  \begin{algorithmic}[1]
  \Procedure{Shifted inverse iteration}{$\Psi_0$, $\eta$, $m$}
    \For{$k=1, \dots, N_\text{IP}$}
    \State Solve $(\opr{H} - \eta) \Psi_{k} = \Psi_{k-1}$ where $\Psi_{k}$  is an MPS with bond dimension $m$
    \State $\Psi_{k} = \Psi_{k}/\norm{\Psi_{k}}$
    \EndFor
  \EndProcedure
  \end{algorithmic}
  \caption{Pseudocode of the DMRG[IP] algorithm.}
  \label{alg:DMRG_IP}
\end{algorithm}
\begin{algorithm}[htbp!]
  \begin{algorithmic}[1]
    \Procedure{DMRG[FEAST]}{$N_\text{p}$, $m$, $N_\text{FEAST}$, $N_\text{GMRES}$, $\Phi_0^{(i)}$}
      \For{$n_\text{iter}$=0, $N_\text{FEAST}$}
        \For{$i$=0, $N$}
            \For{$k$=0, $N_\text{p}$}
                \State Solve Eq.~\ref{eq:linfeast} to calculate $\Phi^{(i,k)}$ as an MPS with bond dimension $m$.
            \EndFor
        \EndFor
        \State Construct the Hamiltonian representation $\bm{H}$ in the $S_N$ basis.
        \begin{equation*}
            H_{ij} = \sum_{k,h}^{N_p} \omega_k^\star \omega_h \langle \Phi^{(i,k)} | \mathcal{H} | \Phi^{(j,h)} \rangle
        \end{equation*}
        \State Construct the overlap matrix $\bm{S}$ for the $S_N$ basis.
        \begin{equation*}
            S_{ij} = \sum_{k,h}^{N_p} \omega_k^\star \omega_h \langle \Phi^{(i,k)} | \Phi^{(j,h)} \rangle
        \end{equation*}
        \State Solve the generalized eigenvalue problem.
        \begin{equation*}
            \bm{H} \bm{V}^{n_\text{iter}} = \bm{E}^{n_\text{iter}} \bm{S} \bm{V}^{n_\text{iter}}
        \end{equation*}
        \State Calculate convergence measure $\eta$ (Eq.~\ref{eq:SecondConvergence})
        \If{$\eta < \eta_\text{FEAST}$}
            \State Exit.
        \Else
            \State Construct the guess for next iteration:
            \For{$i$=0, $N$}
                \State
                \begin{equation*}
                    \Phi_{(n_\text{iter}+1)}^{i} 
                        = \sum_{k=1}^{N_\text{p}} \sum_{j=1}^{N} V_{i,j} \omega_k \Phi^{(j,k)}
                \end{equation*}
            \EndFor
        \EndIf
      \EndFor
    \EndProcedure
  \end{algorithmic}
  \caption{Pseudocode of the DMRG[FEAST] algorithm.}
  \label{alg:DMRG_FEAST}
\end{algorithm}
\noindent To establish the convergence properties and accuracy of DMRG[IP] and DMRG[FEAST], we discuss both methods for the vibrational problem outlined in Section~\ref{subsec:vDMRG}. We refer to the resulting algorithm as vDMRG[IP] and vDMRG[FEAST], respectively. We note that a  DMRG[FEAST] calculation is defined by seven parameters: 1) the bond dimension $m$, 2) the number of iterations of the iterative linear system solver ($N_\text{GMRES}$), 3) the number of microiterations ($N_\text{micro}$), 4) the interval for the complex contour integration $I_E$, 5) the number of quadrature points to approximate the complex contour integral $N_p$,  6) the number of MPS guess wave functions, $N$ and 7) the choice for the initial guess MPS.
We optimize MPS with the single-site variant of DMRG and set $N_\text{GMRES}$=50, unless otherwise specificed.
We rely on a 8-point Gauss-Hermite quadrature for all vDMRG[FEAST] calculations.
Where possible, we compare the vDMRG[IP] and vDMRG[FEAST] results with excited-state energies calculated with the constrained optimization variant of vDMRG (the vDMRG[ortho] method in Ref.~\citenum{Baiardi2019_HighEnergy-vDMRG}).
We refer to this methods simply as ``vDMRG'', for the sake of readability. \\

\noindent A key parameter of both DMRG[FEAST] and DMRG[IP] is the initial guess MPS.
In the following, we adopt two choices.
We will refer to an MPS where the entries $\bm{M}_{a_{i-1},a_i}^{\sigma_i}$ are initialized randomly $\sigma_i < \sigma_i^\text{max}$ 
as ``random guess'' 
($\sigma_i^\text{max} < N_\text{max}$ is a mode-specific parameter that is also initialized randomly).
The other entries are set to 0. The constraint $\sigma_i < \sigma_i^\text{max}$  avoids the generation of guess wave functions corresponding to vibrational states with a large excitation degree that would yield slow convergence rates.
Alternatively, we will refer 
to MPSs constructed from the ONV corresponding to a specific harmonic target state
as ``harmonic guess''.
In all cases, we set $N_\text{max}=6$, \textit{i.e.} we include 6 harmonic basis functions per mode in the MPS wave function.

\noindent We applied vDMRG[IP] and vDMRG[FEAST] to two molecular systems: ethylene and uracil.
Ethylene is a 12-mode molecule that will serve as a reference to validate the new methods introduced in the present work by comparing them to results obtained by us\cite{Baiardi2017_VDMRG,Baiardi2019_HighEnergy-vDMRG} with the excited-state vDMRG variants introduced in Section~\ref{sec:theory}.
Our original work relied on the PES reported in Ref.~\citenum{Delahaye2014_EthylenePES} that we approximated as a sixth-order Taylor expansion with the \texttt{PyPES} software.\cite{Crittenden2015_PyPES}
Here instead, we rely on the PES reported in Ref.~\citenum{Berkelbach2021_HBCI-Vibrational} and compare our results with the corresponding vibrational heat-bath CI (vHBCI) calculations.
Note that in our previous work on the anharmonic vibrational structure of ethylene~\cite{Baiardi2017_VDMRG} we have neglected fifth- and sixth-order force constants smaller than 0.1~cm$^{-1}$, whereas the PES of Ref.~\citenum{Berkelbach2021_HBCI-Vibrational} includes the complete sixth-order force field.
Small deviations with respect to the data reported in Refs.~\citenum{Baiardi2017_VDMRG} and \citenum{Baiardi2019_HighEnergy-vDMRG} are therefore to be expected.

\noindent Uracil will serve as a test case to analyse the efficiency of large-scale vDMRG[IP] and vDMRG[FEAST] calculations, for molecules with more than 20 vibrational modes, which are hardly targeted by alternative full CI-based algorithms.
We rely on the PES reported in Ref.~\citenum{Carrington2018_Uracil}, which is represented as a fourth-order Taylor expansion and includes the full third-order force field, as well as diagonal and semi-diagonal fourth-order force constants.
The harmonic frequencies were calculated with Coupled Cluster,\cite{Puzzarini2011_Uracil-Composite} while third- and fourth-order anharmonic force constants were calculated with second-order M{\o}ller-Plesset perturbation theory.\cite{Krasnoshchekov2015_Uracil}
As already noted in Ref.~\citenum{Carrington2018_Uracil}, this PES diverges to large negative energies for large displacements from the equilibrium position.
This effect was referred to as PES ``holes'' in Ref.~\citenum{Carrington2018_Uracil}.
This divergence affects the accuracy of variational calculations dramatically\cite{Carrington2018_Uracil} since the energy minimization can converge to an unphysical wave function with an energy lower than the ground state.
The effect is particularly relevant for vDMRG where the excitation degree of the full CI wave function is not truncated, which 
leads to the exploration of arbitrarily high-energy configurations.
We have removed all anharmonic couplings between the 4 lowest-energy modes and the X-H stretching vibrations and report the resulting PES in the Supplementary Material.
In order to assign the optimized excited states obtained with vDMRG[FEAST] or vDMRG[IP], we applied the stochastic reconstruction of the CI wave function introduced in Ref.~\citenum{Boguslawski2011_SRCAS} that we generalized to vibrational wave functions in Ref.~\citenum{Baiardi2019_HighEnergy-vDMRG}.
This algorithm samples the probability distribution $p_{\bm{\sigma}}$ associated with the MPS, defined as

\begin{equation}
  p_{\bm{\sigma}} = \frac{\left\| C_{\bm{\sigma}} \right\|^2}{\sum_{\bm{\sigma}} \left\| C_{\bm{\sigma}} \right\|^2}
  \label{eq:CI_Probability}
\end{equation}
by applying the Metropolis-Hastings algorithm.
All reported energies are absolute energies and reported in cm$^{-1}$, if not otherwise specified).

\section{Strengths and limitations of vDMRG[IP]}
\label{sec:IPI}

\subsection{vDMRG[IP] convergence for ground-state calculations}

The key component of both vDMRG[IP] and vDMRG[FEAST] is the solution of linear systems with the DMRG-like algorithm based on the minimization of Eq.~(\ref{eq:residualMinimum}) with the alternating least squares algorithm.
As a first validation step, we analyze the convergence of this algorithm as applied to the vibrational Hamiltonian of ethylene.

\begin{figure}[htbp!]
  \centering
  \includegraphics[width=.6\textwidth]{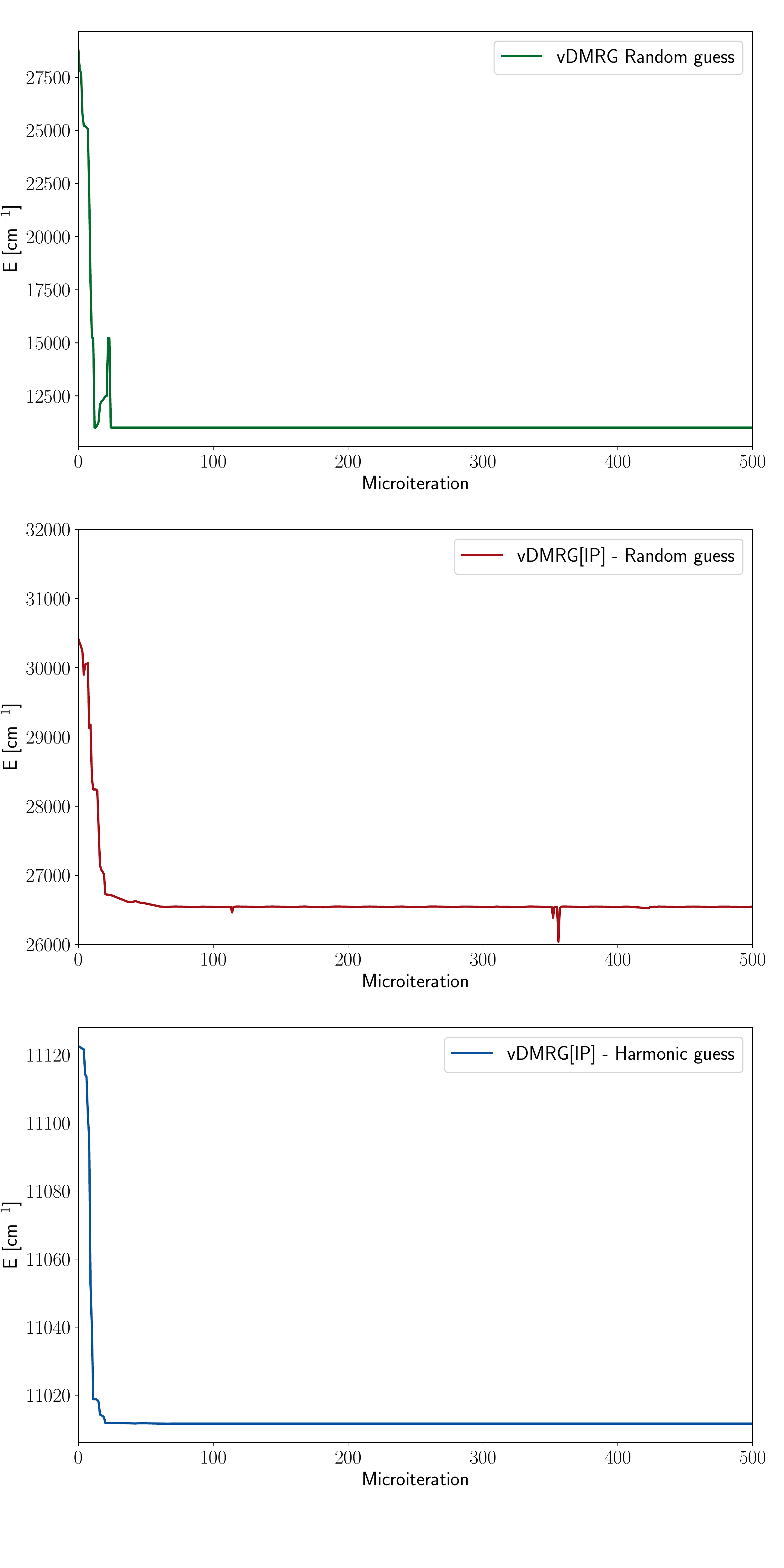}
  \caption{vDMRG (upper panel) and vDMRG[IP] (middle and lower panels) energy convergence for ethylene starting from 
the same random MPS (upper and center panel) and from the harmonic guess associated with the vibrational ground state (lower panel).
  In all cases we set $m$=50 and $N_\text{GMRES}$=50.
  For the vDMRG[IP] calculations, we further set $N_\text{IP}$=1 and $\eta$=11000~cm$^{-1}$.}
  \label{fig:ZPVE_Ethylene}
\end{figure}

\noindent Figure~\ref{fig:ZPVE_Ethylene} depicts the energy convergence of vDMRG[IP] for a single IP iteration ($N_{IP}$=1) such that  the initial guess for the MPS defines both the guess for the left-  and  right-hand side  in Eq.~(\ref{eq:IPI_sol}) for the entire DMRG[IP] run.
We optimize the vibrational ground state by choosing the shift just above the expected zero point vibrational energy 
(ZPVE) of ethylene, $\eta$=11000~cm$^{-1}$.
The initial MPS guess is initialized either randomly or as a harmonic guess.
For the harmonic guess, the energy converges within 24 microiterations that corresponds to 2 sweeps.
The convergence rate is therefore comparable to that of conventional DMRG (upper left panel of Figure~\ref{fig:ZPVE_Ethylene}).
If the calculation is initiated from the random guess, vDMRG[IP] converged in approximately 100 microiteration steps, which corresponds to 10 sweeps.

Note that if the guess is an eigenstate of the S\&I operator $(\mathcal{H} - \eta \mathcal{I})$, the left- and right-hand side of Eq.~(\ref{eq:IPI_sol}) match.
Given the fact that we construct the guess for the solution of the linear system from the right-hand side MPS, the vDMRG[IP] efficiency is in this case highest.
This is the case for the calculations initiated from the harmonic ground-state ONV, but not for these that are started from the default guess.
For this reason, the convergence is slower in the latter case.
Note that, for the harmonic-guess calculation, DMRG[IP] converges to an energy of 11011.61~cm$^{-1}$, in good agreement with the vHBCI reference data.\cite{Berkelbach2021_HBCI-Vibrational}
A single IP macroiteration step is therefore sufficient to converge the ground-state energy of ethylene.
This is not the case for the calculations initiated from the random guess that converges to a higher energy of approximately 26500~cm$^{-1}$.
In this case, multiple applications of the S\&I operator would be required to converge DMRG[IP], since the overlap of the initial MPS and the targeted ground state is small.

\begin{figure}[htbp!]
  \centering
  \includegraphics[width=.9\textwidth]{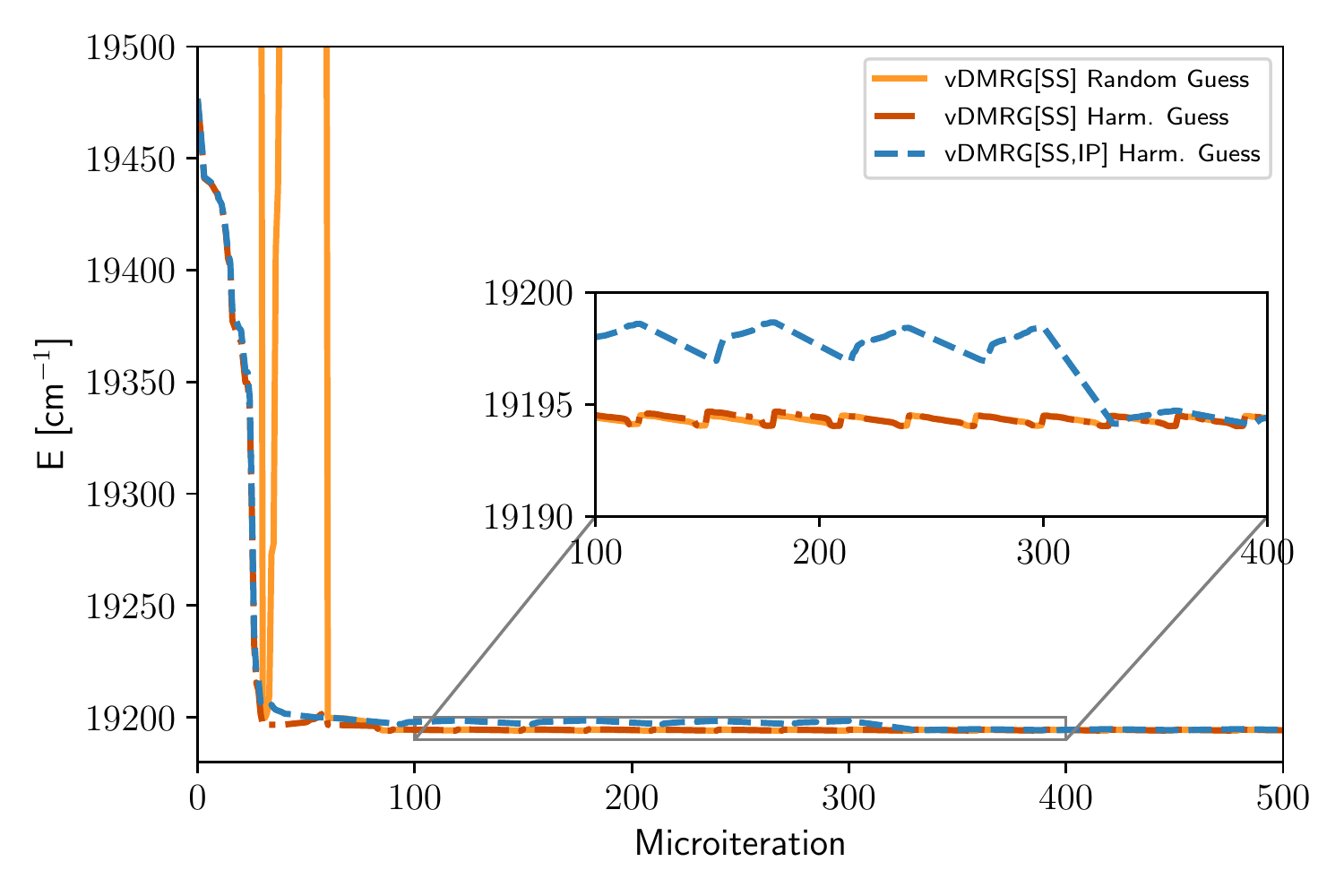}
  \caption{Energy convergence of vDMRG (red lines) and vDMRG[IP] (blue lines) as applied to the optimization of the vibrational ground-state of uracil.
  We set $\eta$=19000~cm$^{-1}$, $m$=50, and $N_\text{IP}$=5 for all vDMRG[IP] calculations.}
  \label{fig:UR_GS_IPIvsDMRG}
\end{figure}

\noindent As we show in Figure~\ref{fig:UR_GS_IPIvsDMRG}, we observed the same trend also for the larger test case, i.e., for uracil.
In this case, we set the shift parameter to 19000~cm$^{-1}$, an energy value that is lower than the converged vDMRG energy of 19193.96~cm$^{-1}$.
Moreover, we set $N_\text{IP}$=5 so that a new IP macroiteration is started after 300 microiterations.
As for ethylene, the energy reaches a plateau after 2 sweeps.
However, the corresponding energy value is 5~cm$^{-1}$ higher than the converged vDMRG energy.
Hence, the solution of the linear system converges after 2 sweeps, but $N_\text{IP}$=1 is no longer
sufficient to converge the overall vDMRG[IP] calculation.

\begin{figure}[htbp!]
  \centering
  \includegraphics[width=.9\textwidth]{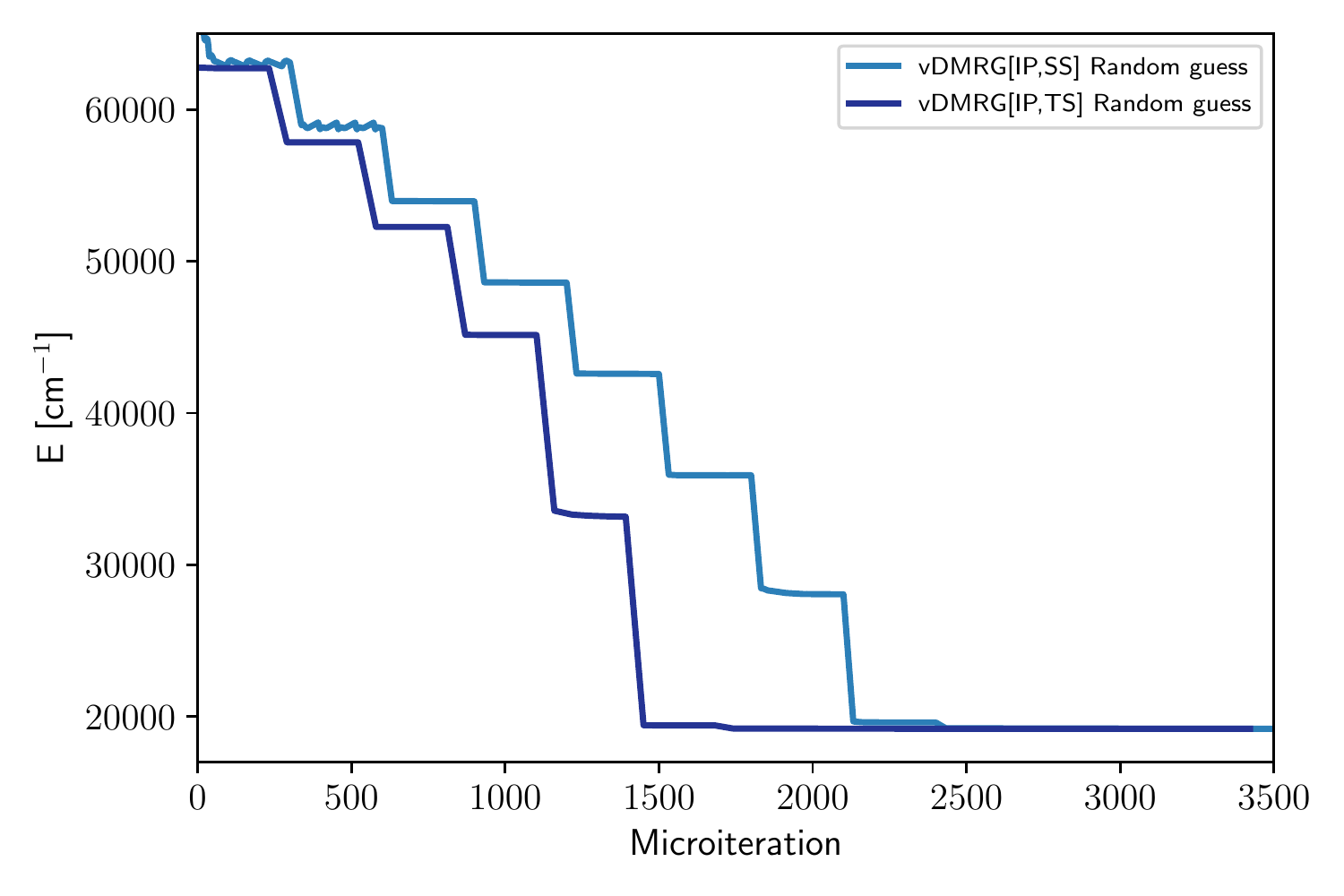}
  \caption{Energy convergence of vDMRG[IP,SS] (dark blue line) and vDMRG[IP,TS] (light blue line) with $\eta$=19000~cm$^{-1}$, $N_\text{IP}$=5, $N_\text{GMRES}=50$, and $m$=50.}
  \label{fig:UR_IPI_Random}
\end{figure}

\noindent The previous analysis does not apply to calculations starting from a random MPS guess.
As we show in Figure~\ref{fig:UR_GS_IPIvsDMRG}, the convergence of vDMRG is almost
unchanged compared to the harmonic guess, while the vDMRG[IP] converges only after 8 macroiterations.
The vDMRG[IP] convergence is slightly enhanced based on the two-site solver that yields convergence after 6 macroiterations.
The number of macroiterations required to converge vDMRG[IP] increases by decreasing the overlap of the initial guess with the final, optimized wave function.
vDMRG[IP] is therefore less efficient than vDMRG for a generic random guess.
We will show later how vDMRG[FEAST] is less dependent on the guess than vDMRG[IP].

\subsection{Excited-state targeting with vDMRG[IP]}

\begin{figure}[htbp!]
  \centering
  \includegraphics[width=.9\textwidth]{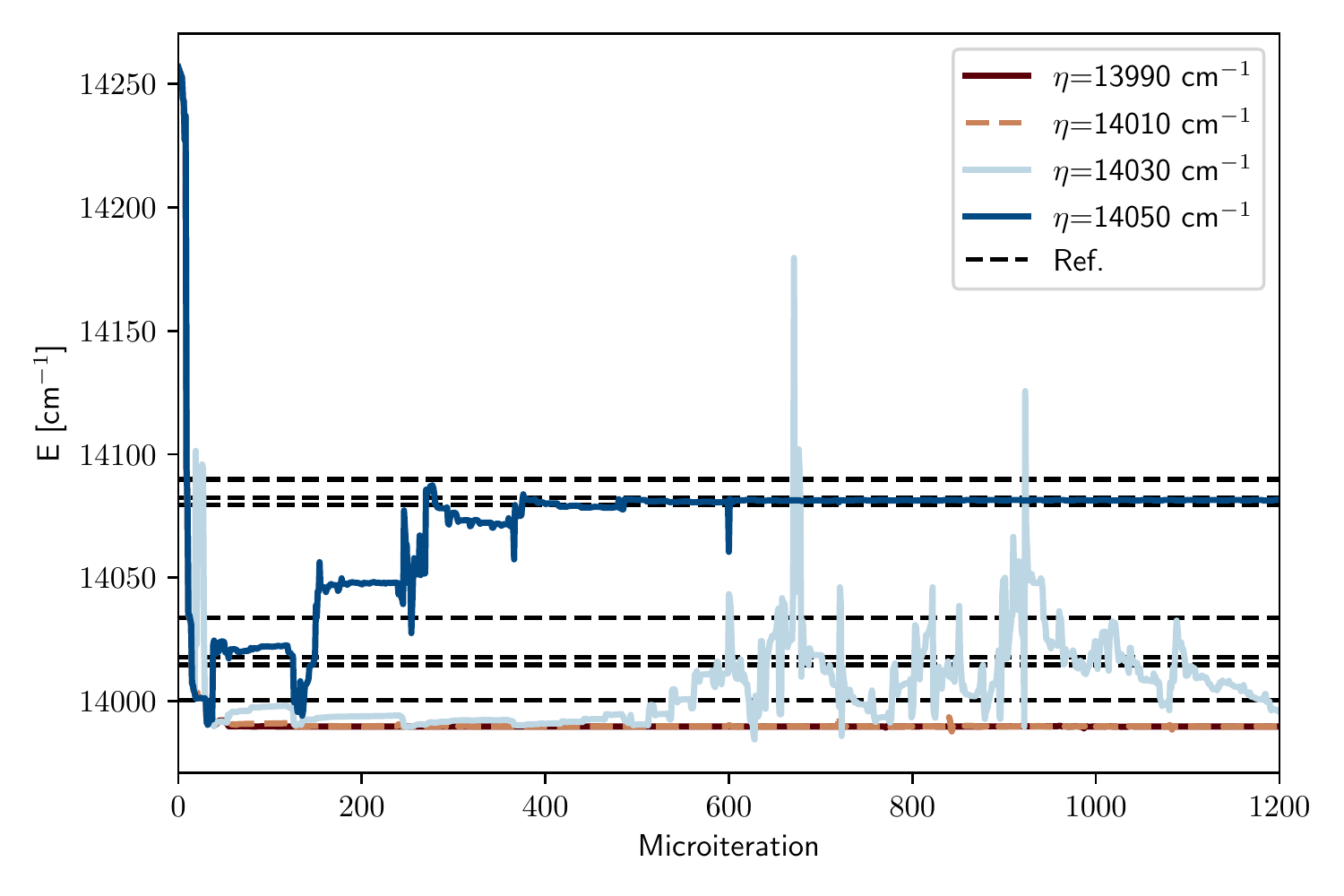}
  \caption{Energy convergence of vDMRG[IP] for ethylene in terms of the number of sweeps obtained with different $\eta$ values, $m$=50, N$_\text{GMRES}$=50, and N$_\text{IP}$=5.
  Selected reference vibrational energies taken from Ref.~\citenum{Delahaye2014_EthylenePES} are represented as horizontal black lines.}
  \label{fig:C2H4_IPI_CHStretching}
\end{figure}

\noindent Although DMRG[IP] can reliably optimize low-energy vibrational states, it becomes completely unfeasible
for excited state calculations as we show in Figure~\ref{fig:C2H4_IPI_CHStretching}, where we optimize the vibrational state associated with the symmetric CH stretching mode of ethylene with vDMRG[IP].
The vDMRG[S\&I] energy of this state calculated based on the fourth-order Taylor-series-expanded PES 
was 13982.3~cm$^{-1}$.\cite{Baiardi2019_HighEnergy-vDMRG}
We therefore chose four shift values $\eta$ ranging from 13980~cm$^{-1}$ to 14050~cm$^{-1}$.
In all cases, we rely on the harmonic guess.
As we show in Figure~\ref{fig:C2H4_IPI_CHStretching}, DMRG[IP] converges within 3 sweeps for $\eta$=13990~cm$^{-1}$ and 14010~cm$^{-1}$
The converged energy is 13989.72~cm$^{-1}$, corresponding to an excitation energy of 2978.08~cm$^{-1}$, in good agreement with the reference VCI energy\cite{Delahaye2014_EthylenePES} of 2985.38~cm$^{-1}$ that is reported in Figure~\ref{fig:C2H4_IPI_CHStretching} as dashed black line.
We note that these VCI energies are calculated for the same PES employed in the present work, but also include vibro-rotational couplings in the Hamiltonian.
Therefore, the reference energy is higher by approximately 10~cm$^{-1}$ than the reference one.
For $\eta$=14030~cm$^{-1}$, vDMRG[IP] does not converge, while for $\eta$=14050~cm$^{-1}$ it converges to a higher energy of 14081.54~cm$^{-1}$, corresponding to an excitation energy of 3069.90~cm$^{-1}$
With the stochastic reconstruction of the CI wave function algorithm\cite{Boguslawski2011_SRCAS,Baiardi2017_VDMRG} we assign this energy to the $\nu_{2}+\nu_{12}$ state.
The corresponding excitation energy is in good agreement with the VCI reference energy\cite{Delahaye2014_EthylenePES} of 3074.92~cm$^{-1}$.
Therefore, as soon as $\eta$=14030~cm$^{-1}$, i.e. when $\eta$ falls halfway between the $\nu_{11}$ and $\nu_{2}+\nu_{12}$ energies, vDMRG[IP] converges the $\nu_{2}+\nu_{12}$ state even though the initial guess is the harmonic $\nu_{11}$ state.
DMRG[IP] convergence is therefore affected by the choice for $\eta$, which is independent of the initial guess MPS.
We note that vDMRG[IP] energy convergence is enhanced by the point group symmetry of ethylene.
In fact, reference VCI data\cite{Delahaye2014_EthylenePES} suggest that 5 states possess an energy higher than that of 
the $\nu_{11}$ state and lower than that of 
the $\nu_{2}+\nu_{12}$ one (we included the corresponding reference energies in Figure~\ref{fig:C2H4_IPI_CHStretching} as dashed black lines).
The vDMRG[IP] optimization scheme does not break the wave function symmetry, and therefore, converges only the states associated with the same irreducible representation as the initial guess MPS, which is the fundamental excitation of the $\nu_{11}$ mode and transforms as B$_{1u}$.
The $\nu_{2}+\nu_{12}$ combination band is the first state with an energy higher than $\nu_{11}$ in symmetry B$_{1u}$.
For this reason, vDMRG[IP] will be unstable starting from $\eta$=14030~cm$^{-1}$.
vDMRG[IP] would be even more sensitive to the choice for the shift$\eta$ for a non-symmetric molecule.

\section{Vibrational ground states with vDMRG[FEAST]}

\noindent We now show that vDMRG[FEAST] can overcome the vDMRG[IP] drawbacks outlined above.
We recall that a DMRG[FEAST] calculation is defined by six parameters: 1) the bond dimension $m$, 2) the number of iterations of the iterative linear system solver ($N_\text{GMRES}$), 3) the number of sweeps for the solution of the linear system ($N_\text{sweep}$), 4) the interval for the complex contour integration $I_E$, 5) the number of quadrature points to approximate the complex contour integral, and 6) the number of guesses.
These parameters determine the extent to which the subspace basis is approximated. 
We first apply vDMRG[FEAST] on the energy interval $I_E=[11000, 11100]$ (we report in the following all energies in cm$^{-1}$ if not otherwise specified) for different $N_\text{GMRES}$ and $m$ values.
This interval includes only the vibrational ground state, and therefore, we set $N_\text{guess}$=1.

\begin{table}[htbp!]
  \centering
  \def\arraystretch{1.5}
  \begin{tabular}{ccc|cccc}
  	\hline \hline
  	              &      $m$                          &       &     20      &      20     &     50     &      50     \\
  	              & N$_\text{GMRES}$                  &       &     20      &      50     &     20     &      50     \\
  	\hline
  	              & \multirow{4}{*}{$N_\text{FEAST}$} &   1   &   11286.51  &   14413.97  &  11326.70  &   11047.72  \\
  	Random        &                                   &   2   &   10677.59  &   10794.03  &  10810.65  &   11011.66  \\
  	guess         &                                   &   3   &   10999.47  &   10992.57  &  10777.01  &   11011.64  \\
  	              &                                   &   4   &   10816.20  &   11105.86  &  10995.05  &   11011.64  \\
  	\hline
  	              & \multirow{4}{*}{$N_\text{FEAST}$} &   1   &   11012.01  &   11012.01  &  11011.64  &   11011.64  \\
  	Harmonic      &                                   &   2   &   11012.01  &   11012.01  &  11011.64  &   11011.64  \\
  	guess         &                                   &   3   &   11012.01  &   11012.01  &  11011.64  &   11011.64  \\
  	              &                                   &   4   &   11012.01  &   11012.01  &  11011.64  &   11011.64  \\ 
  	\hline
  	\hline
  \end{tabular}
  \caption{vDMRG[FEAST] energy convergence with respect to $N_\text{FEAST}$ for $E_\text{min}=11000$~cm$^{-1}$, $E_\text{max}=11100$~cm$^{-1}$, $N_\text{sweep}$=5, and different $m$ and $N_\text{GMRES}$ values.
  All energies are reported in cm$^{-1}$.}
  \label{tab:ZPVE_FEAST_C2H4}
\end{table}

\noindent We show in Table~\ref{tab:ZPVE_FEAST_C2H4} that, starting from the random guess, vDMRG[FEAST] converges only with $m$=50 and $N_\text{GMRES}$=50, whereas convergence is not achieved with lower $m$ and $N_\text{GMRES}$ values.
Conversely, vDMRG[FEAST] converges in a single FEAST iteration when the optimization is initiated from the harmonic guess.
We approximately solve the linear system with fixed $m$, $N_\text{GMRES}$, and $N_\text{sweep}$=5 values,
and therefore, approximate the projector $\mathcal{P}_M$ defined in Eq.~(\ref{eq:FEAST_Projector}).
The impact of this approximation on the vDMRG[FEAST] convergence decreases by increasing the overlap of the guess wave function with the final, exact eigenfunction.
The harmonic guess represents more accurately the target state, and hence, it yields faster vDMRG[FEAST] convergence.
Finally, we note that the converged vDMRG[FEAST] ground-state energy (11011.64~cm$^{-1}$) agrees well with the reference energy calculated with heat-bath CI (11011.61~cm$^{-1}$).\cite{Berkelbach2021_HBCI-Vibrational}
This suggests that the discrepancy with the energy reported in our original vDMRG work\cite{Baiardi2017_VDMRG} is due to the differences in the potential energy surface mentioned in Section~\ref{sec:computationalDetails}.

\begin{figure}[htbp!]
  \centering
  \includegraphics[width=.8\textwidth]{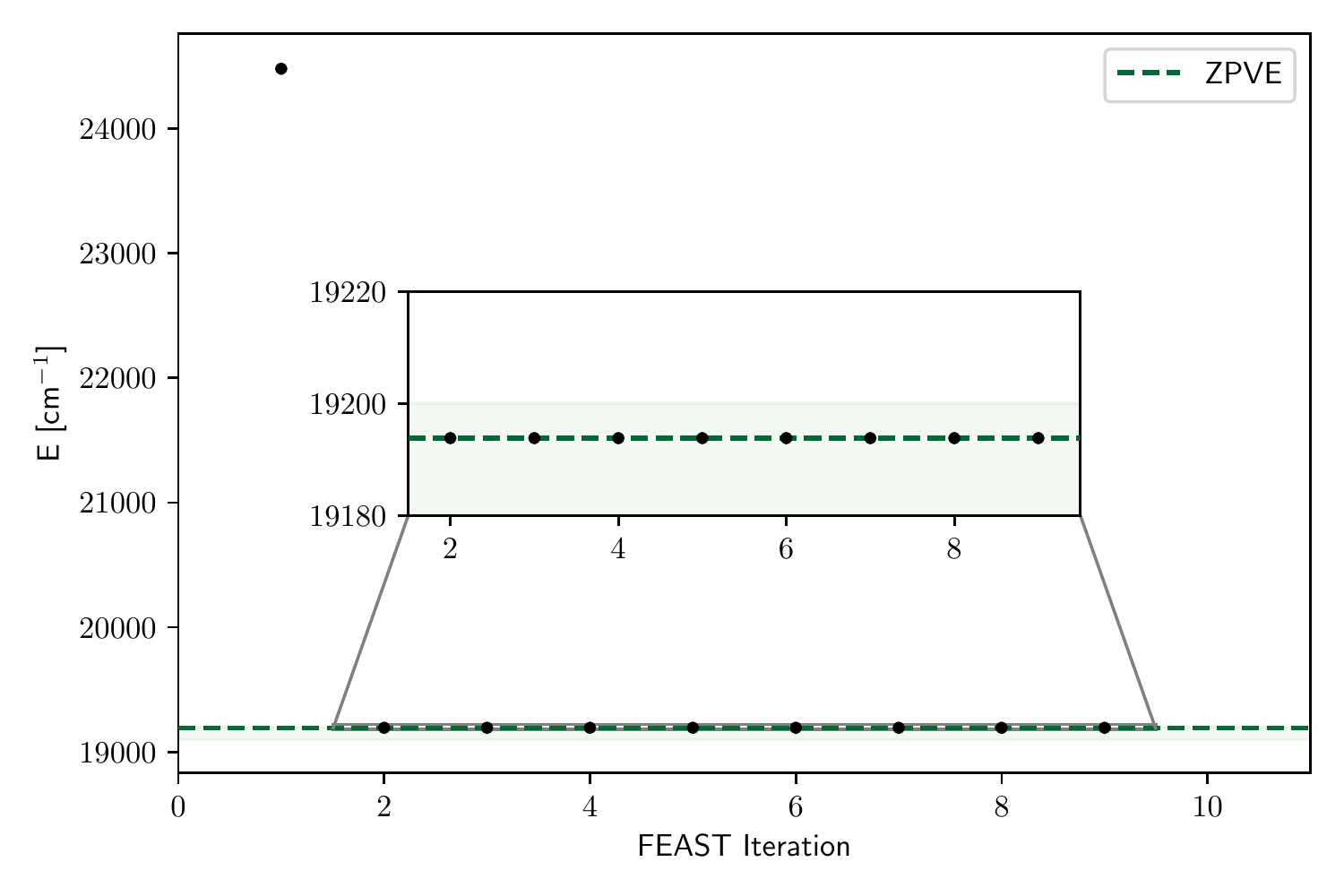}
  \caption{vDMRG[FEAST] energy convergence for the ZPVE of uracil obtained with $E_\text{min}$=19100~cm$^{-1}$, $E_\text{max}$=19200~cm$^{-1}$, $N_\text{GMRES}$=50, $N_\text{sweep}$=5, $m$=50, $N_\text{guess}$, starting from the random guess.
  The reference energy obtained by us for this work with vDMRG is reported as a dashed black line.}
  \label{fig:Uracil_ZPVE_FEAST}
\end{figure}
\noindent We now apply the optimal parameter set, $m$=50 and $N_\text{GMRES}$=50, to optimize the vibrational ground state of uracil.
We report in Figure~\ref{fig:Uracil_ZPVE_FEAST} the vDMRG[FEAST] energy convergence obtained starting from the random guess.
Even though the initial guess does not approximate the final wave function, vDMRG[FEAST] converges in 2 FEAST macroiteration steps, as opposed to vDMRG[IP] that converges, as shown in Figure~\ref{fig:UR_IPI_Random}, in 9 iterations.
This confirms that vDMRG[FEAST] is less sensitive to the choice for the initial guess.

\section{Excited-state optimization with vDMRG[FEAST]}

In this section, we first apply vDMRG[FEAST] to optimize low- and high-energy vibrational states of ethylene to highlight how vDMRG[FEAST] convergence changes with the density of states of the targeted energy interval.
Then, we optimize the low-energy excited states of uracil with vDMRG[FEAST] to demonstrate its reliability for large molecules.

\subsection{vDMRG[FEAST]: optimization of low-lying excited states}

\begin{figure}[htbp!]
  \centering
  \includegraphics[width=.8\textwidth]{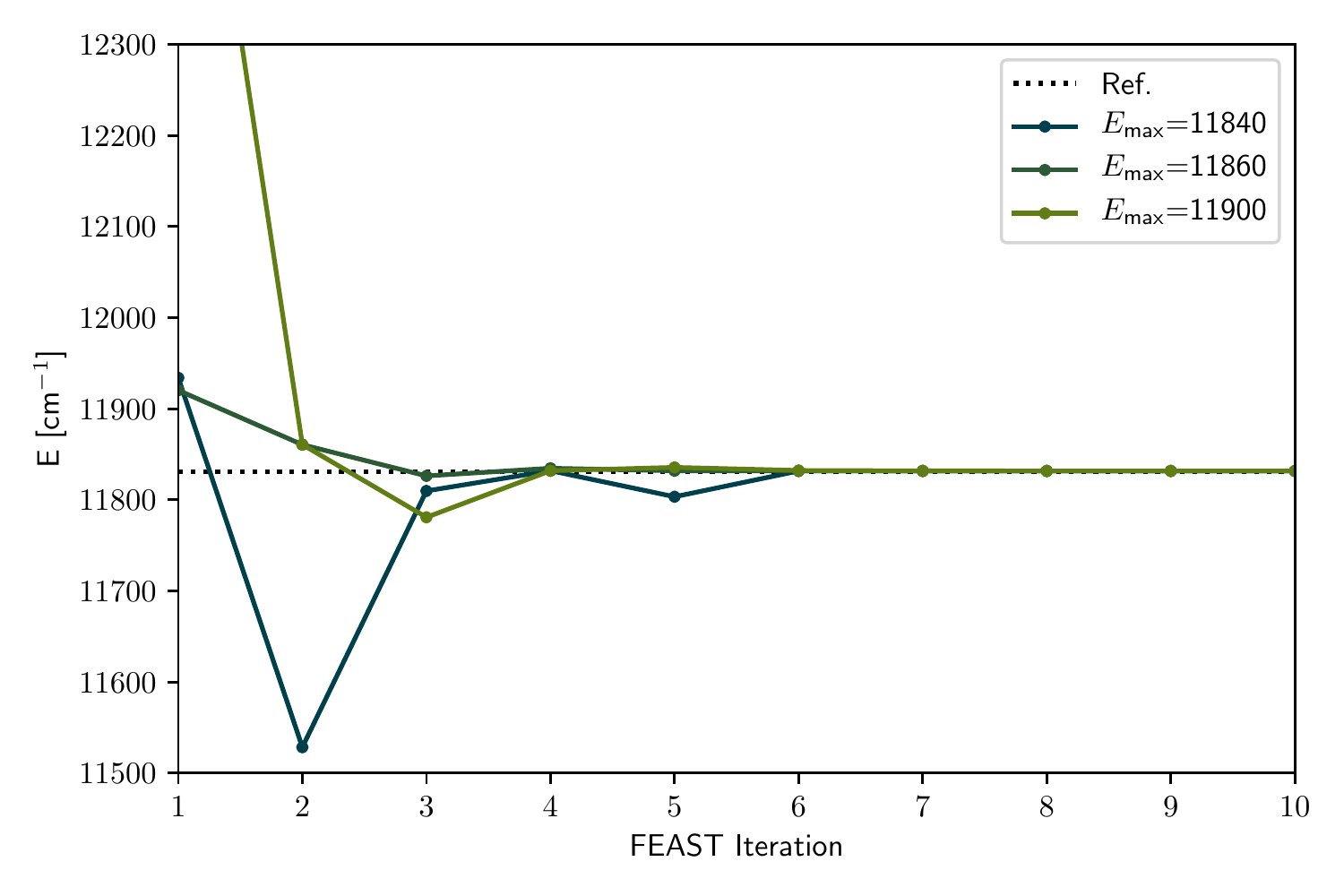}
  \caption{vDMRG[FEAST] energy convergence for the first excited state of ethylene obtained with $N_\text{GMRES}$=50, $N_\text{IP}$=10, $N_\text{guess}$=2, $m$=50, $E_\text{min}$=11820~cm$^{-1}$ and changing $E_\text{max}$ values.
  The reference energy 'Ref.' (dotted black line) was calculated by us for this work with vDMRG[ortho] and is 11831.58~cm$^{-1}$.}
  \label{fig:Ethylene_LowestExcited}
\end{figure}

\noindent Figure~\ref{fig:Ethylene_LowestExcited} shows the vDMRG[FEAST] energy convergence obtained with $E_\text{min}$=11820~cm$^{-1}$ and $E_\text{max}$ values ranging from 11840~cm$^{-1}$ to 11900~cm$^{-1}$ starting from random MPSs.
Reference energies, reported in Table~\ref{tab:LowEnergyExcitedStates_Ethylene} and calculated with vDMRG,\cite{Baiardi2019_HighEnergy-vDMRG} indicate that $I_E$ includes only the first excited vibrational state for all $E_\text{max}$ values.
Even though vDMRG[FEAST] should, in principle, converge that vibrational state already with $N_\text{guess}$=1, we follow Ref.~\citenum{Polizzi2009_FEAST} and set $N_\text{guess}$=2 to enhance algorithm convergence.
In all cases, vDMRG[FEAST] converges in 6 iterations to an energy value that matches the vDMRG reference value (see Table~\ref{tab:LowEnergyExcitedStates_Ethylene}) indicating that the integration interval $I_E$ has only a minor effect on vDMRG[FEAST] convergence.
Moreover, Figure~\ref{fig:Ethylene_LowestExcited} confirms that the linear system can be safely solved by minimizing Eq.~(\ref{eq:residualMinimum_Equivalent}) also for $\eta$ values that are larger than the smaller eigenvalue of $\mathcal{H}_\text{vib}$.

\begin{figure}[htbp!]
  \centering
  \includegraphics[width=.8\textwidth]{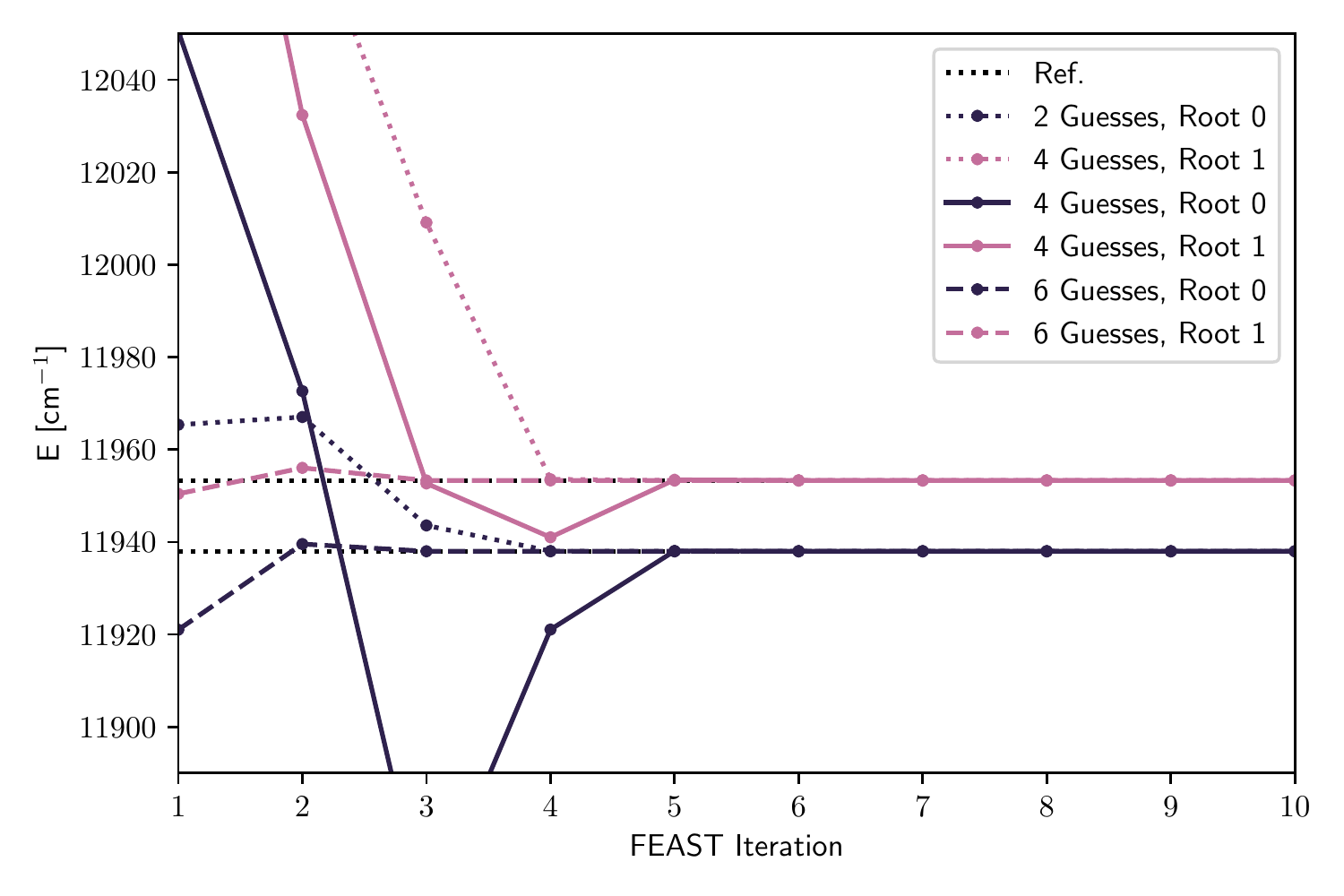}
  \caption{vDMRG[FEAST] energy of the second (purple lines) and third (pink lines) excited states of ethylene obtained with $N_\text{guess}$=2 (dotted lines), $N_\text{guess}$=4 (solid lines), and $N_\text{guess}$=6 (dashed lines).
  In all cases, we set $N_\text{GMRES}$=50, $N_\text{sweep}$=5, $m$=50, $E_\text{min}$=11920~cm$^{-1}$, and $E_\text{min}$=11960~cm$^{-1}$ and started the optimization from random MPSs.
  The reference energies 'Ref.' (dotted black lines) were calculated by us for this work with vDMRG[ortho]
  and $m$=100. They are 11937.92~cm$^{-1}$ and 11953.24~cm$^{-1}$, respectively.}
  \label{fig:Ethylene_SecondLowest}
\end{figure}

\noindent The choice for $N_\text{guess}$ becomes less trivial for energy regions with a higher density of states.
We show in Figure~\ref{fig:Ethylene_SecondLowest} vDMRG[FEAST] energy convergence for $E_\text{min}$=11920~cm$^{-1}$ and $E_\text{max}$=11960~cm$^{-1}$, an energy interval that includes the second and third excited vibrational state (see Table~\ref{tab:LowEnergyExcitedStates_Ethylene}).
Hence, $N_\text{guess}$ must be larger than 2 to converge all states included in this energy inerval.
As shown in Figure~\ref{fig:Ethylene_SecondLowest}, both energies are converged after 5 FEAST iteration steps with $N_\text{guess}$=4 and in 2 FEAST macroiteration steps with $N_\text{guess}$=6.
Therefore, if the initial guess is not an accurate approximation to the target wave function, increasing $N_\text{guess}$ and, hence, the subspace size, will enhance DMRG[FEAST] convergence.

\subsection{Parallel exploration of different energy regions}

\begin{table}[htbp!]
    \centering
    \def\arraystretch{1.5}
    \begin{tabular}{ccc|ccc|cc}
      \hline\hline
          \multirow{2}{*}{E$_\text{min}$} & \multirow{2}{*}{E$_\text{max}$} & \multirow{2}{*}{State} & 
          \multicolumn{3}{c|}{vDMRG[FEAST]} & \multirow{2}{*}{vDMRG} & 
          \multirow{2}{*}{Ref. \citenum{Berkelbach2021_HBCI-Vibrational}}  \\
    &   &   &     $m$=20      &       $m$=50      &      $m$=100      &                 &                \\
      \hline
      \hline
       \multirow{3}{*}{11800} &  \multirow{3}{*}{12000} &
        1   &     11831.93    &     11831.60      &      11831.58     &    11831.58     &    11831.60    \\
        &&
        2   &     11938.23    &     11937.94      &      11937.92     &    11937.92     &    11937.94    \\
        &&
        3   &     11953.56    &     11953.26      &      11953.24     &    11953.24     &    11953.26    \\
      \hline
       \multirow{3}{*}{12000} &  \multirow{3}{*}{12400} &
        4   &     12029.36    &     12029.06      &      12029.05     &    12029.04     &    12029.06    \\
        &&
        5   &     12234.03    &     12233.76      &      12233.71     &    12233.75     &    12233.77    \\
        &&
        6   &     12353.88    &     12353.55      &      12353.53     &    12353.53     &    12353.56    \\
      \hline
       \multirow{5}{*}{12400} &  \multirow{5}{*}{12800} &
        7   &     12450.31    &     12449.93      &      12449.91     &    12449.90     &    12449.92    \\
        &&
        8   &     12634.71    &     12634.05      &      12633.95     &    12633.95     &    12634.39    \\
        &&
        9   &     12666.10    &     12665.00      &      12664.93     &    12664.93     &    12666.82    \\
        &&
       10   &     12760.39    &     12759.67      &      12759.59     &    12759.63     &    12759.66    \\
        &&
       11   &     12778.39    &     12777.81      &      12777.75     &    12777.76     &    12777.80    \\
      \hline\hline
    \end{tabular}
    \caption{vDMRG[FEAST] energy of ethylene obtained for varying $m$ values, $N_\text{guess}$=8, $N_{sweep}$=5, $N_\text{GMRES}$=50 for three energy intervals $I_1=[11800, 12000]$, $I_2=[12000, 12400]$, $I_3=[12400, 12800]$.
    All energies are given in cm$^{-1}$.
    vDMRG reference data were calculated by us for this work with the constrained optimization excited-state vDMRG[ortho] variant. For
    comparison, the last column collects vHBCI data taken from Ref. \citenum{Berkelbach2021_HBCI-Vibrational}.}
    \label{tab:LowEnergyExcitedStates_Ethylene}
\end{table}

\noindent An appealing, key feature of vDMRG[FEAST] is that different energy regions can be explored independently of
one another.
We illustrate this by optimizing the complete set of excited states of ethylene with a vibrational energy lower than 12800~cm$^{-1}$.
We partition the energy spectrum into three regions, $I_1$=[11800~cm$^{-1}$, 12000~cm$^{-1}$], $I_2$=[12000~cm$^{-1}$, 12400~cm$^{-1}$], and $I_3$=[12400~cm$^{-1}$, 12800~cm$^{-1}$], and apply vDMRG[FEAST] to each interval.
In principle, the number of excited states in each energy interval could be estimated based on reference VCI data.\cite{Delahaye2014_EthylenePES} 
However, accurate reference energies are not, in general, available for large molecules that are the ultimate target of vDMRG[FEAST].
Therefore, we set $N_\text{guess}$=8 for all energy intervals, which overestimates the number of states included in each interval, and start the vDMRG[FEAST] optimization from random MPSs.
We show in Table~\ref{tab:LowEnergyExcitedStates_Ethylene} that vDMRG[FEAST] and vDMRG energies agree up to 0.01~cm$^{-1}$ for all states.
Small deviations, in all cases smaller than 1~cm$^{-1}$, are observed compared to reference heat-bath CI energies.\cite{Berkelbach2021_HBCI-Vibrational}
The largest deviation is observed for the 8-th excited state, where the reference energy is higher than the vDMRG[FEAST] one by about 0.44~cm$^{-1}$.
However, the agreement with the vDMRG data suggests that the discrepancy is due to a partial convergence of the vHBCI calculation.

\begin{figure}[htbp!]
  \centering
  \includegraphics[width=.8\textwidth]{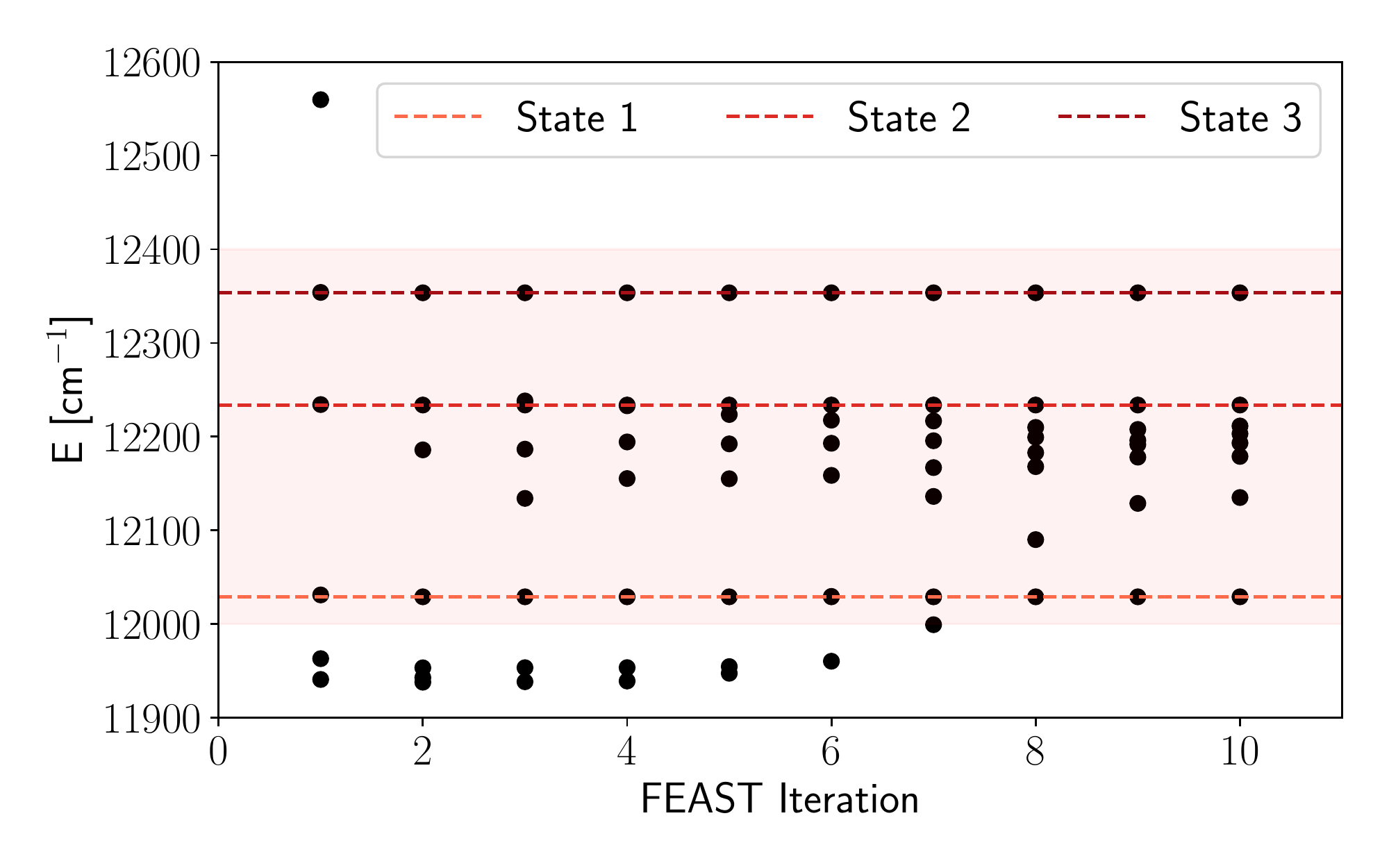}
  \caption{vDMRG[FEAST] energies obtained with $N_\text{GMRES}$=50, $N_\text{sweep}$=50, $m$=50, $N_\text{guess}$=8 and for the energy intervals $I_2$=[12000~cm$^{-1}$, 12400~cm$^{-1}$] (upper panel) and $I_3$=[12400~cm$^{-1}$, 12800~cm$^{-1}$] (lower panel).
  The colored bands represent the energy values included in the $I_2$ and $I_3$ intervals, respectively.
  Reference energies, taken from the vDMRG column of Table~\ref{tab:LowEnergyExcitedStates_Ethylene}, are represented as dashed black lines.}
  \label{fig:Ethylene_ThirdLowest}
\end{figure}

\noindent We analyze in Figure~\ref{fig:Ethylene_ThirdLowest} vDMRG[FEAST] energy convergence for the $I_2$ and $I_3$ energy intervals.
For $I_2$, all three states converge already at the first iteration, even though the optimization is started from randomly initialized MPSs.
Conversely, two iterations are required to converge all states in the $I_3$ interval.
As for the conventional FEAST algorithm,\cite{Polizzi2009_FEAST} also the vDMRG[FEAST] convergence rate increases with the ratio between $N_\text{guess}$ and the number of states included in $I_E$.
Note also that, if $N_\text{guess}$ is larger than the number of states included in $I_E$, some of the roots returned by the subspace diagonalization will correspond to states that do not approximate eigenstates of $\mathcal{H}_\text{vib}$.
This is the case for all MPSs whose energy is not included in the $I_E$ interval.
Then, we calculate the variance for all remaining states based on the algorithm introduced in Ref.~\citenum{Baiardi2019_HighEnergy-vDMRG}
and identify all MPSs with a variance smaller than 100~cm$^{-1}$ as ``physically acceptable'' states.

\subsection{Efficient targeting of high-energy vibrations}
\label{subsec:HighEnergy_FEAST}

\noindent So far, we have applied vDMRG[FEAST] to energy ranges associated with a low density of states.
These intervals can be efficiently targeted also with alternative excited-state vDMRG variants.\cite{Baiardi2019_HighEnergy-vDMRG}
We show now that vDMRG[FEAST] can calculate with the same efficiency also the excitation energy of the four high-energy fundamental transitions of the C-H stretching vibration of ethylene.

\noindent For highly-excited states it is difficult, if not impossible, to select the energy interval $I_E$ so that it includes only a given, target state.
This renders the choice for $I_E$ and $N_\text{guess}$ extremely challenging.
We propose a three-step computational procedure to automate the selection of the $N_\text{guess}$ and $I_E$ parameters in vDMRG[FEAST].
We first obtain an estimate of the energy of the target vibrational state ($E_\text{approx}^\text{target}$), either from the harmonic approximation or cost-efficient approximate anharmonic methods, such as second-order vibrational perturbation theory.\cite{Barone2005_VPT2}
We then apply vDMRG[FEAST] to an interval $I_E^{(1)}$ centered around $E_\text{approx}^\text{target}$.
The convergence of vDMRG[FEAST] is ensured only if $I_E^{(1)}$ includes the energy of the target state ($E_\text{VCI}^\text{target}$), and therefore, if the interval size is larger than the accuracy of the $E_\text{approx}^\text{target}$ estimate.
In this first step we truncate the PES and include only four-body operator in order to limit the MPO size, and hence, the computational cost.
This first vDMRG[FEAST] calculation yields a second, more accurate energy estimate, $E_\text{FEAST,1}^\text{target}$.
Finally, we apply vDMRG[FEAST] on an interval $I_E^{(2)}$ centered around the $E_\text{FEAST,1}^\text{target}$ with the full, six-body potential.
$E_\text{FEAST,1}^\text{target}$ is a much more accurate approximation of the exact energy compared to $E_\text{approx}^\text{target}$ and, therefore $I_E^{(2)}$ should include, ideally, only the target energy.

\begin{figure}[htbp!]
  \centering
  \includegraphics[width=\textwidth]{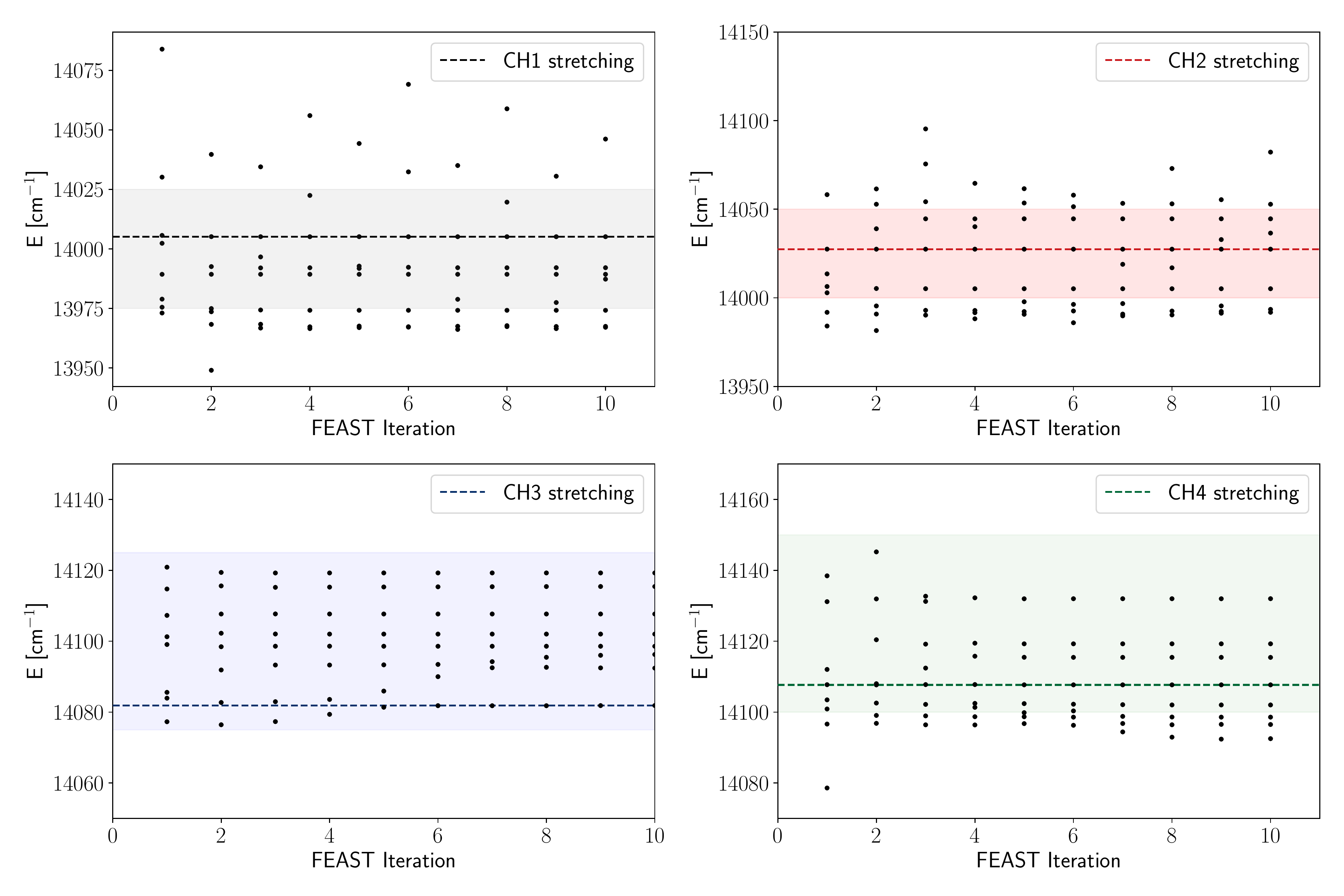}
  \caption{vDMRG[FEAST] energy convergence for the fundamental excitation of the first (upper left panel), second (upper right panel), third (lower left panel), and fourth (lower right panel) C-H stretching mode. 
  We set the energy interval to $E_\text{min}$=13975~cm$^{-1}$ and $E_\text{max}$=14025~cm$^{-1}$ (upper left panel), $E_\text{min}$=14000~cm$^{-1}$ and $E_\text{max}$=14050~cm$^{-1}$ (upper right panel), $E_\text{min}$=14075~cm$^{-1}$ and $E_\text{max}$=14125~cm$^{-1}$ (lower left panel), and $E_\text{min}$=14100~cm$^{-1}$ and $E_\text{max}$=14150~cm$^{-1}$ (lower right panel).
  In all cases, we set $N_\text{GMRES}$=50, $m$=50, $N_\text{quad}$=8, and $N_\text{guess}$=8.
  We report the final, converged energies of the target state with a black dashed line.}
  \label{fig:CH_Quartic_FEAST}
\end{figure}

\noindent Instead of relying on approximate anharmonic vibrational-structure calculation for the first step, we obtain $E_\text{approx}^\text{target}$ from the VCI data of Ref.~\citenum{Delahaye2014_EthylenePES} and set the size of $I_E^{(1)}$ to 50~cm$^{-1}$, to mimic the inaccuracy of the reference data.
Figure~\ref{fig:CH_Quartic_FEAST} exhibits the vDMRG[FEAST] convergence of the first vDMRG[FEAST] calculation phase for the fundamental excitation of the four C-H vibrations (labelled in the following as CH1, CH2, CH3, and CH4, in increasing energy order).
In all cases, we rely on N$_\text{guess}$=8 guess MPSs, the harmonic one and 7 random guesses.
We identified the target state among the N$_\text{guess}$ MPSs returned by DMRG[FEAST] based on the stochastic reconstruction of the CI wave function algorithm introduced above.\cite{Boguslawski2011_SRCAS}
The calculated results are given in Figure~\ref{fig:CH_Quartic_FEAST} and in Table~\ref{tab:C2H4_CH_Convergence}.
Note that, even though $N_\text{guess}$=8 in all cases, the number of final eigenvalues included in the target interval changes with $I_E$ and ranges from 2, for the CH2 mode, to 8, for the CH3 and CH4 modes.
The latter calculations could be repeated with a larger $N_\text{guess}$ value to ensure that all eigenstates included in $I_E$ are converged.
However, this is not necessary because this first calculation step returns only an energy estimate for the second, more accurate vDMRG[FEAST] calculation phase.

\begin{table}
  \centering
  \def\arraystretch{1.5}
  \begin{tabular}{cccccc}
    \hline \hline
        PES      & $N_\text{guess}$ &   CH1   &   CH2   &    CH3    &    CH4    \\
    \hline
      \text{4 body}
                 &      8           &  13989.32  &  14027.44  &   14081.85   &   14107.68   \\
    \hline
      \multirow{2}{*}{6 body}  
                 &      1           &  13989.85  &  14027.75  &   14082.78   &   14103.20   \\
                 &      4           &  13989.64  &  14028.03  &   14082.78   &   14103.10   \\
    \hline
        VCI\cite{Delahaye2014_EthylenePES}    
                 &                  &  14000.29  &  14033.90  &   14094.77   &   14116.16   \\
    \hline \hline
  \end{tabular}
  \caption{vDMRG[FEAST] energies of four C-H stretching modes of ethylene calculated based on the four-body PES and an interval $I_E^{(1)}$ with size 50~cm$^{-1}$, and with the full six-body potential and an interval $I_E^{(2)}$ with size 10~cm$^{-1}$.
  The vDMRG[FEAST] parameters are $N_\text{GMRES}$=50, $m$=50, $N_\text{quad}$=8, and $N_\text{sweep}$=5.
  VCI reference energies taken from Ref. \citenum{Delahaye2014_EthylenePES} are given for comparison.
  All energies are reported in cm$^{-1}$.}
  \label{tab:C2H4_CH_Convergence}
\end{table}

\noindent We run this second set of vDMRG[FEAST] calculations with an interval $I_E$ centered on the energies reported in Table~\ref{tab:C2H4_CH_Convergence}, with $\left| E_\text{max} - E_\text{min} \right| = 10$~cm$^{-1}$ and varying $N_\text{guess}$ values.
As expected, the vDMRG[FEAST] convergence with $N_\text{guess}$ is faster for this smaller energy interval, and a reliable estimate of the vibrational energies is already obtained with $N_\text{guess}$=1 with variations below 1~cm$^{-1}$ obtained with $N_\text{guess}$=4.
Note that the reference VCI energies\cite{Delahaye2014_EthylenePES} are consistently lower compared to the vDMRG[FEAST] results by approximately 10~cm$^{-1}$.
As already noted above, such a discrepancy is to be expected since we neglected vibro-rotational couplings, which are instead included in the VCI calculation of Ref.~\citenum{Delahaye2014_EthylenePES}.

\subsection{Towards large-scale vDMRG[FEAST]: uracil}

\noindent Finally, we demonstrate the reliability of vDMRG[FEAST] for large-scale anharmonic calculations.
We optimize the 11 lowest-energy excited states of our second benchmark molecule, uracil.
Table~\ref{tab:Uracil_ExcitedStates} contains the vDMRG[FEAST] energies obtained with $m$=50 and 100, together with reference data obtained with vDMRG.

\begin{table}
  \begin{tabular}{cc|cc|cc}
	\hline \hline
	$I_E$  &  State  &      $m$=50         &      $m$=100         &  Ref. \citenum{Carrington2018_Uracil} & vDMRG($m$=50) \\
	\hline
	  [19200, 19300]
	       &   ZPVE  &    19218.45         &                      &        18969.21                   &     19193.19  \\
	\hline
	\multirow{3}{*}{[19300, 19500]}
	       &    1    &       146.30        &       147.28         &          140.02                   &      147.58   \\
	       &    2    &       164.49        &       165.50         &          157.09                   &      165.83   \\
	       &    3    &       292.03        &       292.65         &          294.53                   &      293.46   \\
	\hline
	\multirow{3}{*}{[19500, 19600]}
  	      &     4    &       312.87        &       311.98         &          385.15                   &      312.91   \\
	      &     5    &       331.27        &       330.45         &          379.36                   &      331.31   \\
	      &     6    &       386.62        &       386.32         &          510.81                   &      386.53   \\
	\hline
	\multirow{5}{*}{[19600, 19700]}
 	      &     7    &       419.81        &       419.34         &          535.08                   &      419.80  \\
	      &     8    &       437.33        &       435.89         &          543.15                   &      437.34  \\
	      &     9    &       457.97        &       456.28         &          542.77                   &      457.98  \\
	      &    10    &       477.62        &       475.99         &          528.68                   &      520.01  \\
	      &    11    &       496.19        &       494.73         &          651.91                   &      477.64  \\
	\hline
 	\hline
  \end{tabular}
  \caption{vDMRG[FEAST] energies of the 11 lowest-energy vibrational states of uracil calculated with $N_\text{GMRES}$=50, $N_\text{quad}$=8, and varying $I_E$ and $m$ values.
Reference data taken from  Ref. \citenum{Carrington2018_Uracil} are given for comparison.
  All calculations are based on the harmonic guess.}
  \label{tab:Uracil_ExcitedStates}
\end{table}

\noindent Following the same strategy adopted for ethylene, we partition the energy region in three regions, $I_1$=[19300~cm$^{-1}$, 19500~cm$^{-1}$], $I_2$=[19500~cm$^{-1}$, 19600~cm$^{-1}$], and $I_3$=[19600~cm$^{-1}$, 19700~cm$^{-1}$].
In view of the results discussed above, we enhanced the vDMRG[FEAST] convergence by starting the optimization from the harmonic guess in all cases.
Table~\ref{tab:Uracil_ExcitedStates} also provides the energies obtained with vDMRG.
Both vDMRG and vDMRG[FEAST] converge the energy of all vibrational states below 1~cm$^{-1}$ already with $m$=50, therefore confirming that molecular anharmonic vibrational wave functions can be encoded as compact MPSs.
Moreover, even though the vDMRG[FEAST] energies consistently match their vDMRG counterpart, state swappings are observed.
For instance, the energy of the 10-th state is higher than that of the 11-th one.
Root swapping effects largely affect the efficiency of vDMRG, especially for nearly-degenerate high-energy states.
This is not the case of vDMRG[FEAST], however, which calculates each state independently of the others and not in a sequential fashion.

\section{Conclusions}
\label{sec:conclusions}
In this work, we introduced the DMRG[FEAST] algorithm to efficiently optimize excited states of high-dimensional, many-body Hamiltonians, represented as matrix product states.
DMRG[FEAST] applies the FEAST algorithm\cite{Polizzi2009_FEAST} to wave functions and operators encoded as matrix product states and matrix product operators, respectively.\cite{McCulloch2007_FromMPStoDMRG}
We show that DMRG[FEAST] overcomes the main limitations of the existing excited state DMRG algorithms.\cite{Dorando2007_TargetingExcitedStates,Keller2015_MPS-MPO-SQHamiltonian,Devakul2017,Yu2017_ShiftAndInvertMPS,Baiardi2019_HighEnergy-vDMRG}
First, it enables optimizing all the eigenfunctions with an energy lying in a given energy interval without the need of estimating \textit{a priori} the energy of the target excited states.
Moreover, it does not require calculating powers of the Hamiltonian, which are difficult to encode as compact matrix product operators.\cite{Cangiani2013_Folded,Baiardi2019_HighEnergy-vDMRG}
Finally, all DMRG[FEAST] calculation steps are trivially parallelizable, which makes DMRG[FEAST] particularly appealing for large-scale excited-state calculation.

\noindent We applied DMRG[FEAST] to the optimization of high-energy vibrationally excited states of molecular systems with the vibrational DMRG algorithm, although it can be applied to other quantum chemical problems (e.g., in electronic structure
theory) equally well.
Vibrational-structure calculations are, however, the perfect application target for DMRG[FEAST] because calculating vibrational spectra requires obtaining many excitation energies.
We benchmarked the accuracy of DMRG[FEAST] on ethylene against VCI reference data.\cite{Delahaye2014_EthylenePES}
Then, we applied DMRG[FEAST] to calculate the anharmonic vibrational energies of uracil, a system that is a major challenge for state-of-the-art full configuration-interaction methods.\cite{Carrington2018_Uracil}

\noindent Our future work will focus on further enhancing the DMRG[FEAST] generality and efficiency.
Here, we applied vDMRG[FEAST] to optimize specific vibrational states, such as the fundamental excitation energy of the C-H stretching modes of ethylene and the low-energy excited states of uracil.
However, DMRG[FEAST] can be straightforwardly extended to optimize all the excited states in a given energy interval by partitioning it into multiple sub-intervals and applying DMRG[FEAST] to each of them.
Each energy subinterval can be targeted independently from the others, and therefore, the overall algorithm can be trivially parallelized.
Work is currently in progress in our group on automating DMRG[FEAST] based on such a strategy.
The application of DMRG[FEAST] to electronic structure calculations will be particularly appealing for calculating high-energy core-excited states that are difficult to target with conventional multi-configurational methods.
The resulting algorithm would enable the calculation of X-ray spectra beyond the linear response\cite{Nakatani2013_LinearResponse-DMRG} and the state-averaged restricted active space\cite{Lundberg2016_RAS-SCF_Core} approximations.
Finally, we note that DMRG[FEAST] is not tailored to a specific tensor network parameterization and can be extended to the optimization of multi-dimensional tensor network states.\cite{Larsson2019_TTNS-Vibrational}

\begin{acknowledgement}
This work was supported by ETH Z\"{u}rich through the ETH Fellowship No. FEL-49 18-1.
The authors gratefully acknowledge Dr.~Andrea Muolo (Hebrew University of Jerusalem) for fruitful discussions about the FEAST algorithm.
\end{acknowledgement}


\providecommand{\latin}[1]{#1}
\makeatletter
\providecommand{\doi}
  {\begingroup\let\do\@makeother\dospecials
  \catcode`\{=1 \catcode`\}=2 \doi@aux}
\providecommand{\doi@aux}[1]{\endgroup\texttt{#1}}
\makeatother
\providecommand*\mcitethebibliography{\thebibliography}
\csname @ifundefined\endcsname{endmcitethebibliography}
  {\let\endmcitethebibliography\endthebibliography}{}

%

\end{document}